\documentclass[aps,prb,reprint]{revtex4-1}
\usepackage[dvipdfmx]{graphicx}% Include figure files ( add by mk )
\usepackage{dcolumn}% Align table columns on decimal point
\usepackage{bm}% bold math
\usepackage{epsfig}
\usepackage{braket}
\usepackage{amsmath}
\usepackage{here}
\usepackage{amsmath}
\usepackage{amsfonts}
\usepackage{ulem}
\usepackage{xcolor}
\begin{document}
\draft
%%\preprint{HEP/123-qed}
\title{Thermal Hall effect and topological edge states in a square lattice antiferromagnet}
\author{Masataka Kawano and Chisa Hotta}
\address{Department of Basic Science, University of Tokyo, Meguro-ku, Tokyo 153-8902, Japan}
\date{\today}
%%{submitted in May 22, 2002}
\begin{abstract}
We show that the two dimensional spatial inversion-symmetry (SIS) broken square lattice antiferromagnet
with easy-plane spin anisotropy exhibits a thermal Hall effect and 
the edge modes characterized by the $\mathbb{Z}_{2}$ topological invariant. 
These topological properties require a nonzero Berry curvature, 
and its origin is ascribed to the Dzyaloshinskii-Moriya (DM) interactions or the noncoplanar magnetic ordering 
generating a U(1) gauge field that couples to the kinetic motion of magnons. 
Although this picture is established in ferromagnets on the kagome and pyrochlore lattices, 
it does not apply to our square lattice model since such gauge field cancels out in an edge shared geometry. 
Instead, our case has an analogy with the anomalous Hall effect of 
Rashba electronic system where the spin orbit coupling generates an SU(2) gauge field. 
The two species of magnons defined on antiferromagnetic sublattices can be regarded 
as the pseudo-spin degrees of freedom of magnons. 
The DM interactions that emerge due to the SIS breaking serve as a pseudo-spin orbit coupling of magnons 
and generate a Berry curvature, when the direction of the magnetic moments is properly controlled by the magnetic field.
%When the spatial inversion symmetry is broken, 
%the DM interaction becomes antisymmetric about the right and left propagation 
%of magnons that accompanies a pseudo-spin flip, 
%and serves as a pseudo-spin orbit coupling.}
The thermal Hall conductivity of our antiferromagnet shows rapid growth in temperature $T$ beyond the power-law, 
reflecting the almost gapless low energy branch of the antiferromagnet, 
which distinctively differs from the $T^{7/2}$-dependence in the pyrochlore ferromagnets. 
%The topological phase transition occurs by the in-plane rotation of the magnetic field, 
%and the edge modes are protected by the $\mathbb{Z}_{2}$ topological invariant. 
The present system serves as a standard model for the noncentrosymmetric crystals
Ba$_{2}$MnGe$_{2}$O$_{7}$ and Ba$_{2}$CoGe$_{2}$O$_{7}$. 
\end{abstract}
\maketitle
%\begin{multicols}{2}
\narrowtext
%%%%%%%%%%%%%%%%%%%%%%%%%%%%%%%%%%%%%%%%%%%%%%%%%%%%%%%%%%%%%%%%%%%%%%%%%%%%%%%%%%%%%%%%
\section{Introduction}
\label{sec:intro}
Transport phenomena in insulating magnets are intensively pursued these days,  
since it turned out that the magnons, quasi-particles of insulating magnets, 
behave quite similar to the electrons in semiconductors 
and exhibit a thermal Hall effect\cite{review}, 
a magnon spin Nernst effect\cite{cheng16,zyuzin16,shiomi17}, 
and a magnon-driven spin Seebeck effect\cite{jin17}. 
In the typical magnetically ordered insulating states, 
the low energy collective excitation is well-described by the one-body picture of magnon. 
Unlike the electron, the magnon is a charge-neutral quasiparticle which does not feel a Lorentz force. 
Instead, the magnon can feel a Berry curvature, 
which is a fictitious magnetic field in momentum space. 
Accordingly, when driven by a temperature gradient, 
magnons acquire an anomalous velocity and exhibit a thermal Hall effect\cite{onose10,onose12,katsura10,matsumoto11prl,matsumoto11prb,matsumoto14,mook14the,hirschberger15,lee15,owerre17prb,owerre17epl,owerre17com-1,laurell18,laurell17,owerre16-1,owerre16-2,owerre16prb,owerre16app,owerre17com-2,owerre17com-3,owerre17-1,owerre17-2,owerre17app,owerre18,hoogdalem13,nakata17F,nakata17AF,cao15,owerre17star-1,owerre17star-2}.
A nonzero Berry curvature often indicates the topological nature of magnons 
represented by the protected edge modes and Chern numbers, 
which is found in kagome\cite{owerre17prb,owerre17epl,zhang13,mook14topo,chisnell15,mook15-1,mook15-2,mook16kagome,seshadri18}, pyrochlore\cite{li16,mook16pyro,mook17,su17pyro}, and honeycomb lattices\cite{owerre16prb,owerre16app,owerre16-1,owerre16-2,owerre17com-2,owerre17com-3,owerre17-1,owerre17-2,kim16,wang17,owerre17sirep,pantaleon17,owerre18sirep,su17honey,wang18}. 
\par
\begin{figure}[tbp]
\includegraphics[width=6cm]{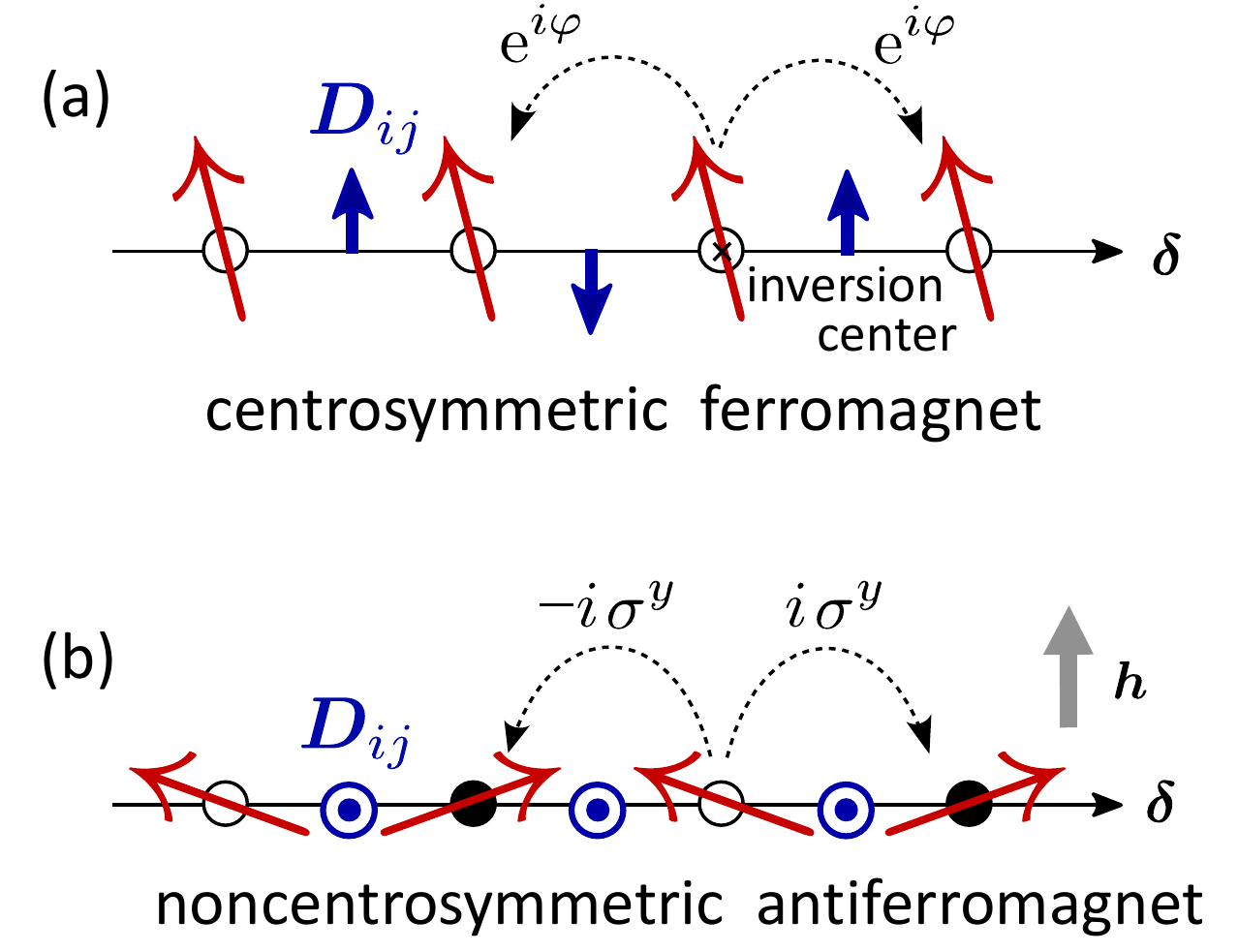}
\caption{Schematic illustration of 
(a) ferromagnet with SIS and (b) antiferromagnet without SIS. 
The DM vector $\bm D_{ij}$ is defined by taking the $i\rightarrow j$ in the $\bm \delta$-direction. 
In (a) the DM vectors align in a staggered manner to keep the SIS, 
and the hopping of a magnon carry a U(1) gauge field or Peierls phase, $\varphi$, which is known to generate a finite Berry curvature. 
In (b) when a magnon hops to its neighbor it virtually flips the pseudo-spin 
(since the magnons on different sublattices carry opposite spins) via $i \sigma_y$. 
Due to the SIS breaking, the sign of hopping to the left and right differs, 
and together with $i \sigma_y$ generates a finite Berry curvature, which is our main proposal. 
}
\label{f0}
\end{figure}

In the previous studies on magnon thermal Hall effects, 
there always exists an effective U(1) gauge field that couples to magnons and gives a finite Berry curvature. 
Two main mechanisms to generate such gauge field are proposed. 
One is to have a noncoplanar magnetic order, on which the magnons hop around, 
feel the curved spin space and acquire a phase\cite{owerre17prb,owerre17star-1,owerre17star-2,laurell18}. 
The other is to make use of an antisymmetric exchange, $\bm D_{ij} \cdot[\bm S_i\times \bm S_j]$, between spins, 
$\bm S_i$ and $\bm S_j$, 
called Dzyaloshinskii-Moriya (DM) interaction\cite{dzyaloshinsky1958,moriya1960}, 
which is finite for bonds that have no inversion center. 
In ferromagnets realized in centrosymmetric crystals, the DM vector $\bm D_{ij}$ 
aligned in a staggered manner along the bonds [see Fig.\ref{f0}(a)]
gives a Peierls phase, i.e., a U(1) gauge field, to the hopping of magnons. 
This happens only when the ordered moments point in the direction {\it parallel or antiparallel} 
to the $\bm D_{ij}$ vectors, which can be controlled by the spin anisotropy and the magnetic field.
\par
However, the presence of DM interaction or a noncoplanar ordering alone is not a sufficient condition to have a finite Berry curvature. 
Indeed, Refs.[\onlinecite{onose12}] and [\onlinecite{katsura10}] show that 
the {\it no-go} condition to have a finite Berry curvature is to have the edge shared lattice geometry, 
whose symmetry combined with the time reversal symmetry cancels out the effect of U(1) gauge field\cite{katsura10,onose12}.  
The particular choices of lattices that are exempt from this no-go condition are the corner shared kagome\cite{katsura10,onose12,mook14the,hirschberger15,lee15,owerre17prb,owerre17epl,owerre17com-1,laurell18} and pyrochlore\cite{onose10,onose12,laurell17} lattices, 
and also the Haldane-type honeycomb lattice \cite{owerre16-1,owerre16-2,owerre16prb,owerre16app,owerre17com-2,owerre17com-3,owerre17-1,owerre17-2,owerre17app,owerre18} with a staggered magnetic field. 
\par
The above mentioned mechanism is discussed in a series of centrosymmetric materials 
having spatial inversion symmetry (SIS), 
from which the square lattice systems were excluded as {\it no-go} examples\cite{honey}.
We show that for {\it the noncentrosymmetric} square lattice antiferromagnets, 
the thermal Hall effect and the topologically protected edge modes can emerge 
from a different mechanism. 
%by properly controlling the direction of the ordered moments by the spin anisotropy and the magnetic field. 
%*%*%*%*%*%*%*%*%*%*%*%*%*%*%*%*%*%*
%*%*%*%*%*%*% fig0 *%*%*%*%*%*%*%*%*
%*%*%*%*%*%*%*%*%*%*%*%*%*%*%*%*%*%*
\par
In the electronic systems, the anomalous Hall effect is found in
\textit{the noncentrosymmetric} materials\cite{jungwirth02,onoda02,fang03,culcer03,jungwirth03,culcer04,yao04,dugaev05}. 
Due to the SIS breaking, the Rashba or Dresselhaus spin orbit coupling appears depending on the crystal symmetry\cite{rashba1960,casella1960,bychkov1984,dresselhaus1955}.
This interaction generates an SU(2) gauge field\cite{hatano07} that couples the flipping of spins with the antisymmetric right and left propagations of electrons, 
and generates a finite Berry curvature when the time reversal symmetry (TRS) of the Hamiltonian is broken. 
%Here, the TRS breaking does not necessarily require an external magnetic field, 
%e.g. the couplings with the localized moments is sufficient. 
\par
This kind of phenomenon is not observed in ferromagnetic insulators since the magnons carry spins that point in the unique direction and do not host up and down spin degrees of freedom like electrons. 
In contrast, in antiferromagnets, there are two magnetic sublattices and accordingly, the two species of magnons are defined on these sublattices, which are regarded as pseudo-spin degrees of freedom. 
We show that when there exists a set of DM vectors that align uniformly along the bonds due to the SIS breaking, 
they serve as a pseudo-spin orbit coupling of magnons, 
and generate a finite Berry curvature. 
To activate such effect, one needs to control the direction of the magnetic moments as shown in Fig. \ref{f0}(b) 
by using the spin anisotropy effect and an external magnetic field; 
the antiferromagnetic moments must be placed in the plane 
{\it perpendicular} to the DM vectors and needs to be canted.
\par
The comparison of Figs.\ref{f0}(a) and \ref{f0}(b) highlights the difference 
between the previously studied ferromagnets and our antiferromagnet. 
In the ferromagnets, the DM vectors are activated when they are {\it parallel} to the ordered moments, 
which is exclusive to our case with DM vectors {\it perpendicular} to the moments. 
Therefore, the no-go theorem discussed in the ferromagnets that forbids the thermal Hall effect of magnons in the square lattice does not apply to our case. 
The details will be discussed in \S. II C and in \S.IV. 
\par
The paper is organized as follows; 
in \S.II, we consider the model of two dimensional (2D) SIS broken antiferromagnet, and 
the conditions to have a Berry curvature.
In \S.III we demonstrate that the system hosts a thermal Hall effect and $\mathbb{Z}_{2}$ topological edge states.
In \S.IV, we revisit the previous cases of ferromagnets in Fig.\ref{f0}(a) 
and summarize the no-go condition to clarify the distinct difference from our findings. 
In \S.V, we summarize our results. 
Our model has correspondence with noncentrosymmetric antiferromagnets 
Ba$_{2}$MnGe$_{2}$O$_{7}$\cite{masuda10,murakawa12,iguchi18} and 
Ba$_{2}$CoGe$_{2}$O$_{7}$\cite{zheludev03,murakawa10,murakawa12}, 
as will be discussed in more detail in the next section.

%
%*%*%*%*%*%*%*%*%*%*%*%*%*%*%*%*%*%*%*%*%*%*%*%*
%*%*%*%*%*%*%*%*%*%*%*%*%*%*%*%*%*%*%*%*%*%*%*%*
\section{formulation}
%
%
%*%*%*%*%*%*%*%*%*%*%*%*%*%*%*%*%*%*
%*%*%*%*%*%*% fig2 *%*%*%*%*%*%*%*%*
\begin{figure}[tbp]
\includegraphics[width=8.5cm]{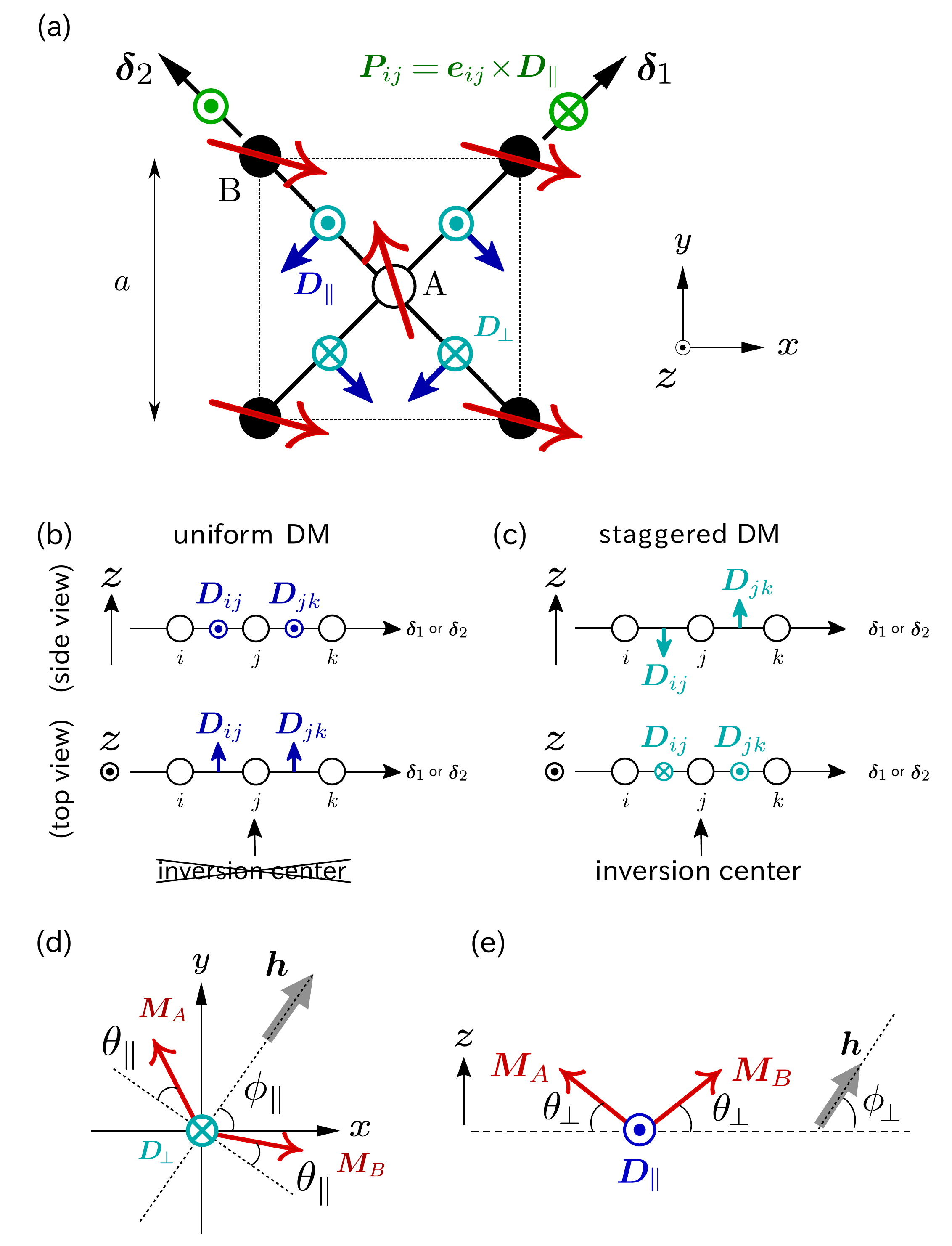}
\caption{2D antiferromagnet with an easy-plane anisotropy, described in Eq.(\ref{ham}). 
(a) Magnetic unit cell in the easy $xy$-plane. 
Open and filled circles represent the A and B magnetic sublattices, respectively, 
and red arrows are the examples of the classical spin configuration in the ground state. 
The DM vector, $\bm D_{ij}$, has the in-plane ($D_\parallel$) and out-of-plane($D_\perp$) elements.
Here, ($i\rightarrow j$) is taken in the $\bm \delta_1$ and $\bm \delta_2$-directions of bonds. 
The green marks pointing perpendicular to the plane are the polarization vector,
$\bm P_{ij}=\bm e_{ij} \times \bm D_\parallel$. 
(b) Uniform and (c) Staggered DM interactions, where only the former breaks the SIS.
The directions of the DM vectors are depicted by running the indices in the $\bm{\delta}_{1}$- or $\bm{\delta}_{2}$-directions.
(d) The angle of the external magnetic field $\bm h$
and (e) the canted angle of the magnetic moments against the direction perpendicular to $\bm h$. }
\label{f1}
\end{figure}
%*%*%*%*%*%*%*%*%*%*%*%*%*%*%*%*%*%*
%*%*%*%*%*%*%*%*%*%*%*%*%*%*%*%*%*%*
%*%*%*%*%*%*%*%*%*%*%*%*%*%*%*%*%*%*%*%*%*%*%*%*
%
\subsection{Model}
We now consider the 2D square lattice antiferromagnet consisting of $N$ spins, whose Hamiltonian reads
\begin{eqnarray}
{\cal H}&=&J\sum_{\langle i,j\rangle}\bm{S}_{i}\cdot\bm{S}_{j}
+\sum_{\langle i,j\rangle}\bm{D}_{ij}\cdot[\bm{S}_{i}\times\bm{S}_{j}] \nonumber \\
&&+\Lambda\sum_{i}(S_{i}^{z})^{2}-\sum_{i}\bm{S}_{i}\cdot\bm{h}, 
\label{ham}
\end{eqnarray}
where $\Lambda >0$ indicates the uniaxial easy-plane magnetic anisotropy, 
and $\bm{D}_{ij}$ is the DM vector defined on the bond connecting the $i$th and $j$th spins. 
The magnetic field $\bm h$ works as a Zeeman term. 
\par
Figure~\ref{f1}(a) shows the magnetic unit cell of our square lattice where open and filled circles represent the A- and B-sublattices, respectively. 
Here, we take the $xy$-axes in the direction rotated by $\pi/4$ from the bond direction for later convenience. 
We define two different vectors, $\bm{\delta}_{1}=(\bm{e}_{x}+\bm{e}_{y})a/2$ and 
$\bm{\delta}_{2}=(-\bm{e}_{x}+\bm{e}_{y})a/2$, along the bonds of the square lattice,
where we take the length of the unit cell as $a=1$.
\par
The DM vector has the in-plane and out-of-plane components, $\bm{D}_{\parallel}$ and $\bm{D}_{\perp}$, 
which keep the $\bar C_4$ symmetry of the lattice. 
$\bm{D}_{\parallel}$ is allowed by the Moriya's rule in the $D_{2d}$ or $T_d$ point group symmetry of the crystal,
which has relevance to the semiconductors with Dresselhaus type spin orbit interaction\cite{dresselhaus1955}. 
Figure~\ref{f1}(a) shows the $\bm D_{ij}$-vectors 
whose indices $i \rightarrow j$ run in the $\bm \delta_1$- and $\bm \delta_2$-directions. 
\par
There are two different ways of aligning the DM vector, 
as shown in Figs.~\ref{f1}(b) and \ref{f1}(c). 
The former breaks the SIS and is called the uniform DM interaction.
The latter staggered DM interaction does not contribute to the SIS breaking of the lattice.
This can be understood simply as follows;
When taking site $j$ as an inversion center,
the inversion operation exchanges $\bm{S}_{i}\leftrightarrow\bm{S}_{k}$.
Therefore, the following part of the Hamiltonian transforms as
\begin{align}
&\bm{D}_{ij}\cdot[\bm{S}_{i}\times\bm{S}_{j}]+\bm{D}_{jk}\cdot[\bm{S}_{j}\times\bm{S}_{k}] \nonumber \\
&\hspace{20pt}\rightarrow-\bm{D}_{jk}\cdot[\bm{S}_{i}\times\bm{S}_{j}]-\bm{D}_{ij}\cdot[\bm{S}_{j}\times\bm{S}_{k}]
.
\end{align}
This straightforwardly indicates that the inversion symmetry is represented by $\bm{D}_{ij}=-\bm{D}_{jk}$,
which holds only in the staggered case [Fig.\ref{f1}(c)].
In our case, $\bm D_{\parallel}$ is a uniform DM interaction and 
$\bm D_{\perp}$ is a staggered DM interaction. 
We see shortly that even though we include $\bm D_\perp$ in the present model, it does not play any dominant role. 
\par
It is convenient to introduce a polarization vector, $\bm P_{ij}= \bm e_{ij} \times \bm D_\parallel$, 
to classify the spatial alignment of $\bm D_{ij}$ vectors, where $\bm e_{ij}$ points to either $\bm \delta_1$ or $\bm \delta_2$. 
The $\bm{P}_{ij}$ along $\bm{\delta}_{1}$ and $\bm{\delta}_{2}$ point in the opposite direction, 
which we call {\it the Dresselhaus type DM interaction}. 
One can define another model by converting $\bm{D}_\parallel$ along the $\bm{\delta}_{2}$-direction up side down\cite{kawano18}. 
In this case, all the $\bm P_{ij}$'s are in the -$z$ direction, which we call {\it the Rashba type DM interaction}, 
since it shares the same geometry with the Rashba spin orbit coupling generated by the electric field perpendicular to the plane\cite{rashba1960,casella1960,bychkov1984}. 
Since the two types of geometry only change the sign of the Hall conductivity, 
we focus on the Dresselhaus type in the present paper. 
The mechanism to realize a nonvanishing Berry curvature is the same for both types.
\par
The Hamiltonian Eq.(\ref{ham}) serves as a standard model for Ba$_{2}$MnGe$_{2}$O$_{7}$ 
and Ba$_{2}$CoGe$_{2}$O$_{7}$ with space group $P\bar{4}2_{1}m$\cite{murakawa12}. 
The MnO$_{4}$ or CoO$_{4}$ tetrahedra form a square lattice, which has an easy-plane anisotropy. 
The slightly canted antiferromagnetic order is realized in the easy-plane below 
$4$K in Ba$_{2}$MnGe$_{2}$O$_{7}$\cite{masuda10} and below $7$K in Ba$_{2}$CoGe$_{2}$O$_{7}$\cite{zheludev03}. 
The canting is induced by the out-of-plane components of the DM interaction, $\bm D_\perp$. 
The magnetic anisotropy inside the easy-plane is tiny in both materials and can be neglected. 
The Dresselhaus type DM interaction, $\bm D_\parallel$, 
is compatible with the crystal symmetry of these materials and should be present, 
although its amplitude is not clarified. 
\par
For later convenience, we decompose the magnetic field, $\bm{h}$,
into the ones parallel and perpendicular to the easy-plane as,
$\bm{h}=\bm{h}_\parallel+\bm{h}_\perp$. 
The in-plane magnetic field, $\bm{h}_{\parallel}$,
fixes the in-plane direction of the magnetic moments [Fig.\ref{f1}(d)].
The out-of-plane magnetic field, $\bm{h}_{\perp}$,
cants the spins off the plane, as shown in Fig.\ref{f1}(e). 
The angles of these moments are described by the field angle about the Cartesian coordinates, $\phi_\parallel$ and $\phi_\perp$, 
as well as by the canting angles, $\theta_\parallel$ and $\theta_\perp$, against the axes perpendicular to $\bm h$. 
%
%*%*%*%*%*%*%*%*%*%*%*%*%*%*%*%*
%*%*%*%*%*%*%*%*%*%*%*%*%*%*%*%*
%*%*%*%*%*%*%*%*%*%*%*%*%*%*%*%*
\subsection{Spin-wave analysis}
\label{sec:sw}
In the following, we apply a spin-wave analysis to Eq.(\ref{ham}).
For sufficiently small $D_{\parallel}$ compared to $\Lambda$,
the canted antiferromagnetic order is realized (see Appendix \ref{app1}), 
whose angles $\theta_{\parallel}$ and $\theta_{\perp}$ are determined numerically by minimizing the classical energy,
\begin{align}
E(\theta_{\parallel},\theta_{\perp})&=
-2JNS^{2}(\mathrm{cos}2\theta_{\parallel}\mathrm{cos}^{2}\theta_{\perp}-\mathrm{sin}^{2}\theta_{\perp}) \nonumber \\
&-2D_{\perp}NS^{2}\mathrm{sin}2\theta_{\parallel}\mathrm{cos}^{2}\theta_{\perp}
+\Lambda NS^{2}\mathrm{sin}^{2}\theta_{\perp} \nonumber \\
&-hNS(\mathrm{sin}\phi_{\perp}\mathrm{sin}\theta_{\perp}+\mathrm{cos}\phi_{\perp}\mathrm{sin}\theta_{\parallel}\mathrm{cos}\theta_{\perp}).
\end{align}
Before applying the Holstein-Primakoff transformation, 
we need to set the $z$-axis of each spin independently to these directions 
by the local unitary transformation\cite{zhitomirsky1997}, $U^\dagger{\cal H} U$, with $U$ given as, 
\begin{align}
U=&\mathrm{exp}\Big(i\sum_{j}\frac{\pi}{2}S_{j}^{x}\Big)\mathrm{exp}\Big(-i\sum_{j}(\theta_{\parallel}\mathrm{e}^{i\bm{Q}\cdot\bm{r}_{j}}-\phi_{\parallel})S_{j}^{y}\Big) \nonumber \\
&\times\mathrm{exp}\Big(-i\sum_{j}\theta_{\perp}\mathrm{e}^{i\bm{Q}\cdot\bm{r}_{j}}S_{j}^{x}\Big),
\label{eq:unitary}
\end{align}
where, $\bm{r}_{j}$, is the spatial coordinate of site-$j$, and $\bm{Q}$ is the ordering wave vector which satisfy $\mathrm{e}^{i\bm{Q}\cdot\bm{r}_{j}}=\pm1$ for sublattices A and B, respectively.
The transformed spin operators are antiparallel, and are given by the standard Holstein-Primakoff transformation\cite{holstein1940} for the 
two magnetic sublattices, A and B as, 
\begin{equation}
S_{i}^{+}\simeq\sqrt{2S}a_{i} \hspace{10pt} S_{i}^{-}\simeq\sqrt{2S}a_{i}^{\dagger} \hspace{10pt} S_{i}^{z}=S-a_{i}^{\dagger}a_{i}, 
\label{hp-trans}
\end{equation}
and
\begin{equation}
S_{j}^{+}\simeq\sqrt{2S}b_{j}^{\dagger} \hspace{10pt} S_{j}^{-}\simeq\sqrt{2S}\ b_{j} \hspace{10pt} S_{j}^{z}=-S+b_{j}^{\dagger}b_{j}, 
\end{equation}
respectively.
After a Fourier transformation,
$a_{i}=\sqrt{2/N}\sum_{\bm{k}}a_{\bm{k}}\mathrm{e}^{i\bm{k}\cdot\bm{r}_{i}}$
and
$b_{j}=\sqrt{2/N}\sum_{\bm{k}}b_{\bm{k}}\mathrm{e}^{i\bm{k}\cdot\bm{r}_{j}}$,
we find the bosonic Bogoliubov-de Gennes (BdG) Hamiltonian described 
in the 4$\times$4 form
with a vector $\Phi_{\bm{k}}=(a_{\bm{k}},b_{\bm{k}},a_{-\bm{k}}^{\dagger},b_{-\bm{k}}^{\dagger})^{T}$, as
\begin{equation}
U^{\dagger}{\cal H}U
\simeq\mathrm{const}+\frac{1}{2}\sum_{\bm{k}}\Phi_{\bm{k}}^{\dagger}H_{\mathrm{BdG}}(\bm{k})\Phi_{\bm{k}}, 
\label{eq:uhu}
\end{equation}
\vspace{-20pt}
\begin{equation}
H_{\mathrm{BdG}}(\bm{k})=\left(
\begin{array}{ll}
\Xi_{\bm{k}} & \Delta_{\bm{k}} \\
\Delta_{-\bm{k}}^{*} & \Xi_{-\bm{k}}^{*}
\end{array}
\right),
\label{eq:hk}
\end{equation}
where $\bm k=(k_x,k_y)$ is defined on a 2D reciprocal space.
$\Xi_{\bm{k}}$ and $\Delta_{\bm{k}}$ are the $2\times2$ matrices,
and satisfy $\Xi_{\bm{k}}^{\dagger}=\Xi_{\bm{k}}$ and $\Delta_{\bm{k}}^{\dagger}=\Delta_{-\bm{k}}^{*}$, respectively.
To diagonalize the bosonic BdG Hamiltonian, $H_{\mathrm{BdG}}(\bm{k})$,
we need to solve the following eigenvalue equation\cite{colpa1978},
\begin{equation}
\Sigma^{z}H_{\mathrm{BdG}}(\bm{k})\bm{t}_{\pm}(\bm{k})=\omega_{\pm}(\bm{k})\bm{t}_{\pm}(\bm{k}),
\label{eq:egeq}
\end{equation}
where $\Sigma^{z}=\tau^{z}\otimes\sigma^{0}$, 
and $\tau^{\mu}$, $\sigma^{\mu}$ ($\mu=0,x,y,z$) 
are the Pauli matrix (and unit matrix $\mu=0$) 
acting on a particle-hole space and a sublattice space, respectively.
The magnon bands, $\omega_{\pm}(\bm{k})\geq0$,
and the corresponding eigenvectors, $\bm{t}_{\pm}(\bm{k})=(u_{\pm,A}(\bm{k}), u_{\pm,B}(\bm{k}), v_{\pm,A}(\bm{k}), v_{\pm,B}(\bm{k}))^{T}$, are obtained,
which satisfy the following paraunitary condition,
\begin{equation}
T(\bm{k})=\left(
\begin{array}{llll}
\bm{t}_{+}(\bm{k}) & \bm{t}_{-}(\bm{k}) & \Sigma^{x}\bm{t}_{+}^{*}(-\bm{k}) & \Sigma^{x}\bm{t}_{-}^{*}(-\bm{k})
\end{array}
\right),
\label{eq:para}
\end{equation}
\vspace{-20pt}
\begin{equation}
T^{\dagger}(\bm{k})\Sigma^{z}T(\bm{k})=T(\bm{k})\Sigma^{z}T^{\dagger}(\bm{k})=\Sigma^{z},
\end{equation}
where $\Sigma^{x}=\tau^{x}\otimes\sigma^{0}$.
The index, $\pm$, denotes the upper and lower magnon bands, respectively.
The analytical form of the magnon bands and the eigenvectors are shown in Appendix. \ref{app2}.
%
%*%*%*%*%*%*%*%*%*%*%*%*%*%*%*%*%*%*%*%*%*%*%*%*
%*%*%*%*%*%*%*%*%*%*%*%*%*%*%*%*%*%*%*%*%*%*%*%*
%*%*%*%*%*%*%*%*%*%*%*%*%*%*%*%*
%*%*%*%*%*%*%*%*%*%*%*%*%*%*%*%*%*%*
%*%*%*%*%*%*% fig3 *%*%*%*%*%*%*%*%*
\begin{figure}[tbp]
\includegraphics[width=8cm]{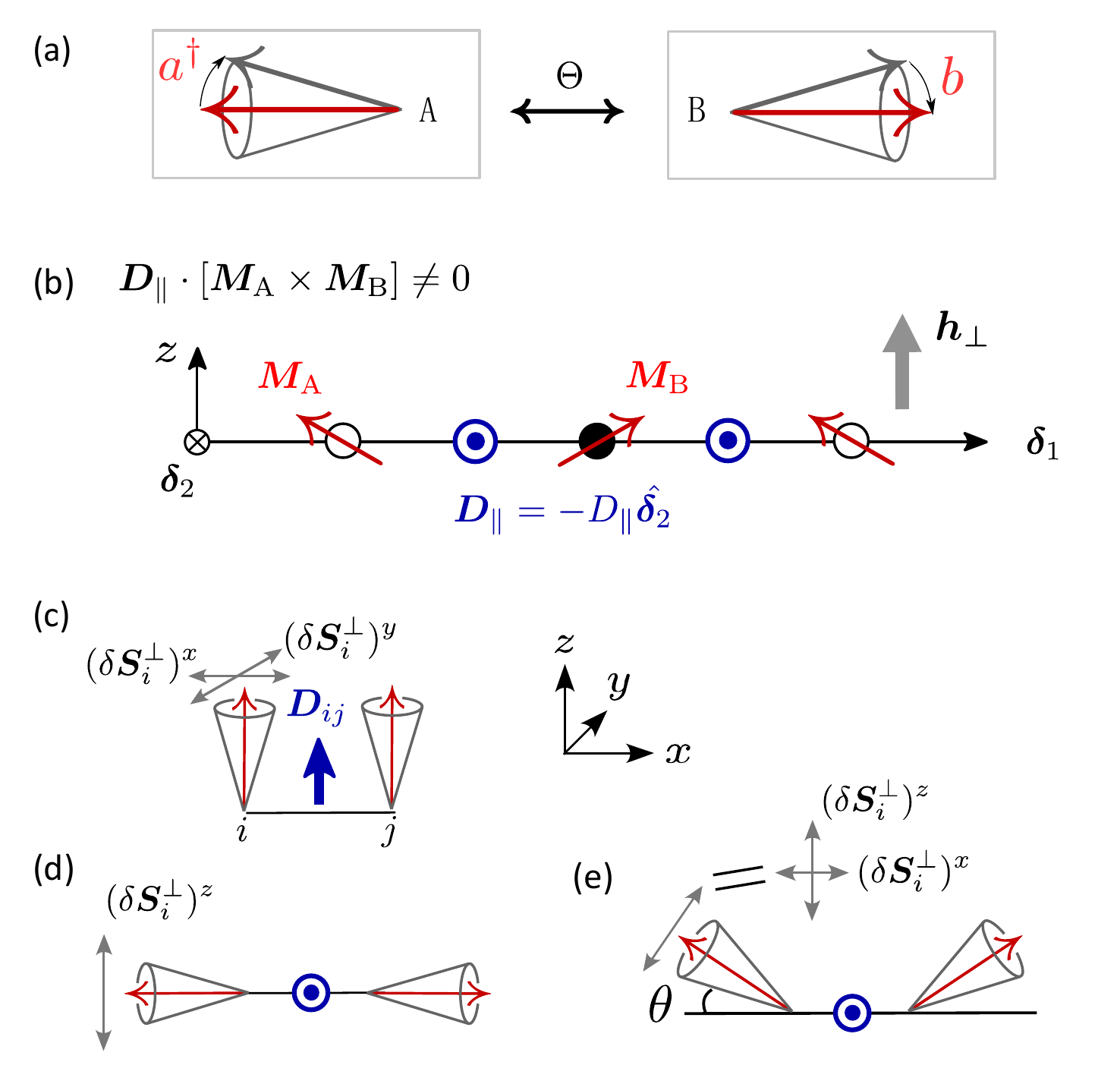}
\caption{(a) Sketch of the effective TRS operation in the present system in Eq.(\ref{eq:ptrs}).
(b) Schematic description of the antiferromagnetic state that gives $\Omega_{xy}^{(\pm)}(\bm k) \ne 0$, 
extracting the $\bm \delta_1$-direction from Fig.\ref{f1}(a). 
When $\bm{h}_{\perp}\neq0$, the ordered magnetic moment and the uniform DM vector form a finite solid angle.
(c)-(e) Schematic illustration of the magnetic moments and DM vector.
The DM vector perpendicular to the collinear magnetic moments as in (d) does not contribute the magnon excitation.
However, when the spins cant as in (e),
the DM vector can couple to the kinetic motion of magnons,
and generates a finite Berry curvature.}
\label{f2}
\end{figure}
%*%*%*%*%*%*%*%*%*%*%*%*%*%*%*%*%*%*
\subsection{Symmetry and Berry curvature}
\label{sec:symm}
A Berry curvature contributes to the thermal Hall effect and the topological nature of the system, 
and is defined as\cite{matsumoto14}
\begin{equation}
\Omega_{xy}^{(\pm)}(\bm{k})=-2\mathrm{Im}\left[\frac{\partial\bm{t}_{\pm}^{\dagger}(\bm{k})}{\partial k_{x}}\Sigma^{z}\frac{\partial\bm{t}_{\pm}(\bm{k})}{\partial k_{y}}\right].
\label{eq:berry}
\end{equation}
The thermal Hall conductivity is given by the following formula\cite{matsumoto14}, 
\begin{equation}
\kappa_{xy}=-\frac{k_{B}^{2}T}{\hbar}\int_{\mathrm{BZ}}\frac{d^{2}\bm{k}}{(2\pi)^{2}}\sum_{\eta=\pm}\left\{c_{2}[f(\omega_{\eta}(\bm{k}))]-\frac{\pi^{2}}{3}\right\}\Omega_{xy}^{(\eta)}(\bm{k}), 
\label{eq:Thermal-Hall}
\end{equation}
where $f(\omega)=1/\{\mathrm{exp}(\omega/k_{B}T)-1\}$ is the Bose distribution function, 
and $c_{2}[x]=\int_{0}^{x}dt[\mathrm{ln}\{(1+t)/t\}]^{2}$. 
\par
The way to understand the condition for a nonvanishing Berry curvature is three-fold. 
First, we see that the A- and B-sublattices form a pseudo-spin degrees of freedom and 
breaking its pseudo-TRS and SIS via $\bm D_\parallel\ne 0$ 
is the necessary condition to have $\Omega_{xy}^{(\pm)}(\bm{k})\ne 0$. 
Second, we need to properly set the relative angles between adjacent magnetic moments and 
the DM vector to have a contribution to the magnon excitation from each bond via $\bm D_{ij}\cdot[\bm S_i\times\bm S_j]$. 
This will require $\bm M_A, \bm M_B\perp \bm D_\parallel$ as well as a canting of the magnetic moments by $\bm h$. 
Finally, we need to have ${\rm Im}(H_{\rm BdG}(\bm{k}))\ne 0$ after summing up the above contributions from bonds, avoiding the cancellations from the symmetry of the system, 
which is attained for the uniform DM vectors in Fig.\ref{f1}(b). 
Once these three conditions are fulfilled, 
$\bm D_{ij}=\bm D_\parallel \ne 0$ serves as a pseudo-spin orbit interaction on magnons, 
and $\Omega_{xy}^{(\pm)}(\bm{k})\ne 0$ is realized in a similar way to the case of the anomalous Hall effect in Rashba electronic systems. 
\par
Often, the symmetry of the system puts some constraints on its Berry curvature. 
In the electronic systems, the SIS gives $\Omega(-\bm{k})=\Omega(\bm{k})$ and 
TRS gives $\Omega(-\bm{k})=-\Omega(\bm{k})$.
The combination of SIS and TRS leads to $\Omega(\bm{k})=0$ for all $\bm{k}$. 
The similar context holds for magnons. 
Let us show that if $\bm{D}_{\parallel}=\bm{0}$, the SIS and TRS in the language of magnons 
are present, and we find $\Omega_{xy}^{(\pm)}(\bm{k})=0$.
For $\bm{D}_{\parallel}=\bm{0}$ the bosonic BdG Hamiltonian in Eq.(\ref{eq:hk}) satisfies the following two equations,
\begin{align}
H_{\mathrm{BdG}}(-\bm{k})=H_{\mathrm{BdG}}(\bm{k}),
\label{eq:sis} \\
\Theta H_{\mathrm{BdG}}(-\bm{k})\Theta^{-1}=H_{\mathrm{BdG}}(\bm{k}), 
\label{eq:ptrs}
\end{align}
where we introduce the anti-unitary operator $\Theta=(\tau^{0}\otimes\sigma^{x})K$, 
using the complex conjugation operator $K$. 
Equation (\ref{eq:sis}) reflects the SIS of the Hamiltonian. 
In Eq.(\ref{eq:ptrs}), the bosonic BdG Hamiltonian is invariant under the symmetry operation represented by $\Theta$.
This operator exchanges $a^{\dagger}\leftrightarrow b$ and $b^{\dagger}\leftrightarrow a$,
namely, the creation of spin moments on A-sublattice is converted to the annihilation of spin moments on B-sublattice pointing in the opposite direction. 
Let us regard the antiferromagnetic moments on two sublattices as a spinor degrees of freedom. 
Then, $\Theta$ serves as {\it an effective time reversal symmetry operator} 
that flips the spinor upside down, as shown schematically in Fig.\ref{f2}(a).
In analogy with the above mentioned electronic systems, 
the combination of SIS in Eqs.(\ref{eq:sis}) and the pseudo-TRS in (\ref{eq:ptrs})
straightforwardly gives $\Omega_{xy}^{(\pm)}(\bm{k})=0$. 
(See \S.\ref{sec:discuss} for the details of the formulation). 
Therefore, $\bm D_\parallel \ne 0$ is the necessary condition to have a nonvanishing Berry curvature. 
\par
To examine the second condition,
we decompose $\bm{S}_{i}$ into the ordered moment and the fluctuation part as, 
$\bm{S}_{i}=\braket{\bm{S}_{i}}+\delta\bm{S}_{i}^{\parallel}+\delta\bm{S}_{i}^{\perp}$,
where $\delta\bm{S}_{i}^{\parallel}\parallel\braket{\bm{S}_{i}}$ and $\delta\bm{S}_{i}^{\perp}\perp\braket{\bm{S}_{i}}$.
$\delta\bm{S}_{i}^{\perp}$ represents the kinetic motion of magnons.
Then, the DM interaction term can be written as,
\begin{equation}
\bm{D}_{ij}\cdot[\bm{S}_{i}\times\bm{S}_{j}]\simeq\bm{D}_{ij}\cdot[\delta\bm{S}_{i}^{\perp}\times\delta\bm{S}_{j}^{\perp}]+\mathrm{const}.
\label{eq:DM}
\end{equation}
We ignore the on-site terms such as $\bm{D}_{ij}\cdot[\delta\bm{S}_{i}^{\parallel}\times\braket{\bm{S}_{j}}]$,
which do not affect the following discussion.
The first term of the right-hand side of Eq. (\ref{eq:DM}) is nonzero only 
when the fluctuation components of the two spins, $\delta\bm{S}_{i}^{\perp}$ and $\delta\bm{S}_{j}^{\perp}$, 
are neither parallel nor antiparallel and have components perpendicular to $\bm{D}_{ij}$. 
This can be easily attained in a typical ferromagnet shown in Fig.\ref{f2}(c), 
where each spin has the 2D fluctuation $(\delta\bm{S}_{i}^{\perp})^x$ and $(\delta\bm{S}_{i}^{\perp})^y$, 
that can couple to $\bm D_{ij}$ {\it parallel to} the magnetic moments. 
In contrast, for the collinear antiferromagnets in Fig.\ref{f2}(d), 
there is only one element of fluctuation perpendicular to $\bm D_{ij}$ 
as $\delta\bm{S}_{i}^{\perp}\simeq\sqrt{S/2}(0,a_{i}^{\dagger}+a_{i},-i(a_{i}^{\dagger}-a_{i}))$
and $\delta\bm{S}_{j}^{\perp}\simeq\sqrt{S/2}(0,b_{j}+b_{j}^{\dagger},-i(b_{j}-b_{j}^{\dagger}))$, 
which suppresses the coupling as
\begin{equation}
\bm{D}_{ij}\cdot[\delta\bm{S}_{i}^{\perp}\times\delta\bm{S}_{j}^{\perp}]\simeq 0.
\label{eq:DM=0}
\end{equation}
To activate the coupling, we need to 
make $\bm M_A$ and $\bm M_B$ noncollinear 
by applying a magnetic field as shown in Fig.\ref{f2}(e), 
and the $x$- and $z$-components of fluctuation emerge from the canting angle $\theta_\perp \ne 0$ as
\begin{eqnarray}
\delta\bm{S}_{i}^{\perp}&&\simeq
\sqrt{\frac{S}{2}}(-i(a_{i}^{\dagger}-a_{i})\mathrm{sin}\theta_\perp, a_{i}^{\dagger}+a_{i},-i(a_{i}^{\dagger}-a_{i})\mathrm{cos}\theta_\perp), 
\nonumber \\
\delta\bm{S}_{j}^{\perp}&&\simeq
\sqrt{\frac{S}{2}}(i(b_{j}-b_{j}^{\dagger})\mathrm{sin}\theta_\perp, b_{j}+b_{j}^{\dagger},-i(b_{j}-b_{j}^{\dagger})\mathrm{cos}\theta_\perp). 
\end{eqnarray}
Then, by setting $\bm M_A$ and $\bm M_B$ {\it perpendicular} to $\bm{D}_{ij}$ we get the maximum contribution as
%\begin{align}
%\bm{D}_{ij}\cdot[\delta\bm{S}_{i}^{\perp}\times\delta\bm{S}_{j}^{\perp}]\simeq&
%\frac{1}{2}D_{ij}S\mathrm{sin}2\theta_\perp(a_{i}^{\dagger}b_{j}+a_{i}b_{j}^{\dagger}
%\nonumber \\
%&\hspace{15pt}
%-a_{i}^{\dagger}b_{j}^{\dagger}-a_{i}b_{j}).
%\label{eq:cant-DM}
%\end{align}
\begin{equation}
\bm{D}_{ij}\cdot[\delta\bm{S}_{i}^{\perp}\times\delta\bm{S}_{j}^{\perp}]\simeq
\frac{1}{2}D_{ij}S\mathrm{sin}2\theta_\perp(a_{i}^{\dagger}b_{j}-a_{i}^{\dagger}b_{j}^{\dagger}+\mathrm{h.c.}).
\label{eq:cant-DM}
\end{equation}
We see shortly that the thermal Hall effect appears only when applying a field 
and having $\theta_\perp \ne 0$. 
\par
The final goal is to have a finite imaginary contribution to the BdG Hamiltonian 
after summing up Eq.(\ref{eq:cant-DM}) throughout the system. 
If ${\rm Im}\big(H_{\rm BdG}(\bm k)\big)=0$, one can always choose a real eigenvector $\bm t_\pm(k)$, 
which immediately gives $\Omega_{xy}^{(\pm)}(\bm{k})=0$ [see Eq.(\ref{eq:berry})]. 
Therefore, having a finite imaginary part of $H_{\rm BdG}(\bm k)$ is necessary for $\Omega_{xy}^{(\pm)}(\bm{k})\ne 0$. 
Notice that Fig.\ref{f2}(e) does not generate a U(1) gauge field that gave such imaginary part 
to the Hamiltonian in ferromagnets, since the coefficient in the right-hand side of Eq.(\ref{eq:cant-DM}) is real.
As mentioned, we are considering $\bm D_{ij}=\bm D_\parallel$ which breaks the SIS, 
namely the hopping of magnon to left and right lattice sites
have different signs as 
\begin{align}
&\frac{1}{2}D_{\parallel}S\mathrm{sin}2\theta_\perp
(a_{j}^{\dagger}b_{j+1}+a_{j}b_{j+1}^{\dagger}
-a_{j}^{\dagger}b_{j-1}+a_{j}b_{j-1}^{\dagger})
\nonumber \\
&\hspace{15pt}
= i D_\parallel \sin 2\theta_\perp\sin k 
a_{k}^{\dagger}b_{k} + {\rm h.c.} 
\nonumber \\
&\hspace{15pt} \propto \psi_k^\dagger \big(\sigma^y\sin k \big)\psi_k, 
\label{eq:rashba-so}
\end{align}
where we introduced $\psi_k=(a_k, b_k)^T$ for convenience as a pseudo-spinor operator. 
Since the hopping converts the $a_j$-magnons to $b_{j\pm 1}$-magnons, it accompanies a pseudo-spin flip, 
and since it is antisymmetric, we find $\pm i \sigma_y$ as we showed in Fig.\ref{f0}(b), 
which is the origin of ${\rm Im}\big(H_{\rm BdG}(\bm k)\big)\ne 0$. 
This has the same form as the well-known Rashba spin orbit term 
of the electronic Hamiltonian for which we set 
$\psi_k=(c_{k\uparrow}, c_{k\downarrow})^T$ 
with $c_{k\sigma}$ an annihilation operator of electron with spin $\sigma$. 
Note that even when Eq.(\ref{eq:cant-DM}) holds, 
if the SIS is not broken and 
the alignment of DM vectors is staggered as in Fig.\ref{f2}(c) 
the sign of the hopping of the left and right moving magnons is the same 
and ${\rm Im}\big(H_{\rm BdG}(\bm k)\big)=0$.
\par
To summarize, $\bm D_\parallel\ne 0$ that breaks the SIS 
serves as an effective Rashba-type spin orbit coupling of the 
magnons whose two sublattices forms a pseudo spin degrees of freedom. 
It also breaks the pseudo-TRS, and $\Omega_{xy}^{(\pm)}(\bm{k}) \ne 0$ becomes feasible. 
On the top of that, we need to cant the antiferromagnetic moments off the collinear configuration by the magnetic field, $\bm{h}_{\perp}\neq\bm{0}$. 
Then, the following term appears in $H_{\mathrm{BdG}}(\bm{k})$,
\begin{equation}
(\tau^{0}\otimes\sigma^{y}-\tau^{x}\otimes\sigma^{y})S\sum_{\mu=1,2}\sum_{\nu\neq\mu}D_{\parallel}\hat{\bm{\delta}}_{\mu}\cdot[\bm{M}_{\mathrm{A}}\times\bm{M}_{\mathrm{B}}]\mathrm{sin}\bm{k}\cdot\bm{\delta}_{\nu},
\label{eq:complex}
\end{equation}
where $\hat{\bm{\delta}}_{\mu}$ is a unit vector pointing in the $\bm{\delta}_{\mu}$-direction ($\mu=1,2$).
This term is a complex matrix 
that originates from the solid angle subtended by 
$\bm{M}_{\mathrm{A}}$, $\bm{M}_{\mathrm{B}}$, and $\bm{D}_{\parallel}$ [see Fig.\ref{f2}(b)], 
and generates a finite Berry curvature.
%*%*%*%*%*%*%*%*
\subsection{$\mathbb{Z}_{2}$ topological invariant}
\label{sec:z2}
\par
A finite Berry curvature may also imply the emergence of a topological phase. 
To examine it, we prepare a square shaped lattice by taking a periodic boundary in the $\bm{\delta}_{1}$-directions, 
extend the $\bm{\delta}_{2}$-direction over $2L$ lattice sites and leave its edges open. 
The Fourier transform is given only for the periodic boundary condition where the wave number, $k_{\bm{\delta}_1}$, is well defined. 
The Hamiltonian yields $2L$ pairs of particle and hole energy bands and we examine whether there are topologically protected edge modes that cross the energy gap.
\par 
The existence of the topologically protected edge modes can be explained using the topological invariant. 
Our system has a band degeneracy at some points in the Brillouin zone so we cannot straightforwardly define a magnon Chern number as in Ref.[\onlinecite{shindou13}]. 
However, a proper choice of the direction of the magnetic field allows the magnon bands to be gapped throughout the Brillouin zone except at the corner, 
and a winding number is well defined along $\bm{k}=\bm k_{\bm{\delta}_{2}}$, parallel to $\bm{\delta}_{2}$.
\par
Practically, the $\mathbb{Z}_2$ topological invariant of the magnon system in one dimension has not been discussed so far. 
We thus define it by extending the one for topological superconductors\cite{sato09,sato17}.
Using the paraunitary matrix in Eq.(\ref{eq:para}),
we first define a gauge field as
\begin{equation}
\bm{A}(\bm{k})=-i\mathrm{Tr}[\Sigma^{z}T^{\dagger}(\bm{k})\Sigma^{z}\bm{\nabla}_{\bm{k}}T(\bm{k})]. 
\end{equation}
which is rewritten using $T^{-1}(\bm{k})=\Sigma^{z}T^{\dagger}(\bm{k})\Sigma^{z}$ as
\begin{equation}
\bm{A}(\bm{k})=\bm{\nabla}_{\bm{k}}\mathrm{arg}\{\mathrm{det}T(\bm{k})\}
\label{eq:A}
\end{equation}
Then, the winding number at each fixed $k_{\bm{\delta}_{1}}$ is constructed as
\begin{eqnarray}
W(k_{\bm{\delta}_{1}})&=&\frac{1}{2}\int_{\mathrm{BZ}(k_{\bm{\delta}_{1}})}dk_{\bm{\delta}_{2}}A_{\bm{\delta}_{2}}(\bm{k})
\nonumber\\
&=& \frac{1}{2}\int_{\mathrm{BZ}(k_{\bm{\delta}_{1}})}dk_{\bm{\delta}_{2}}\frac{\partial}{\partial k_{\bm{\delta}_{2}}}\mathrm{arg}\{\mathrm{det}T(\bm{k})\},
\label{eq:wind}
\end{eqnarray}
where $\mathrm{BZ}(k_{\bm{\delta}_{1}})$ is the line over the Brillouin zone at fixed $k_{\bm{\delta}_{1}}$.
In addition, the following relation holds when $\phi_{\parallel}\neq0$ (mod $\pi/2$), $\bm{D}_{\parallel}\neq\bm{0}$, and $\bm{h}_{\perp}\neq\bm{0}$:
\begin{equation}
\mathrm{det}T(\pm\pi,q)=\mathrm{det}T(q,\pm\pi) \hspace{20pt} (\ q\in\partial\mathrm{BZ}\ ),
\end{equation}
where $\partial\mathrm{BZ}$ is the surface of the Brillouin zone.
The $\mathbb{Z}_{2}$ topological invariant is defined as, 
\begin{equation}
\nu(k_{\bm{\delta}_{1}})=\mathrm{e}^{iW(k_{\bm{\delta}_{1}})}. 
\label{eq:nu}
\end{equation}
The value, $\nu(k_{\bm{\delta}_{1}})=-1$,
discriminates the $\mathbb{Z}_{2}$ topological band structure from the trivial one having $\nu(k_{\bm{\delta}_{1}})=1$.
\par
When $\bm{D}_{\parallel}=\bm{0}$,
there is always a band degeneracy over the Brillouin zone boundary.
Thus, we cannot define the winding number.
When $\bm{h}_{\perp}=\bm{0}$, 
$H_{\mathrm{BdG}}(\bm{k})$ is the real matrix,
and the topological invariant is always trivial,
$\nu(k_{\bm{\delta}_{1}})=1$.
Thus, one can conclude that $\bm{D}_{\parallel}\neq\bm{0}$ and $\bm{h}_{\perp}\neq\bm{0}$ 
are the necessary conditions to have a feasible $\nu(k_{\bm{\delta}_{1}})=-1$.
We will also discuss the $\phi_{\parallel}$ dependence of a topological invariant in \S.\ref{sec:topo}.
%*%*%*%*%*%*%*%*%*%*%*%*%*%*%*%*%*%*
%*%*%*%*%*%*% fig4 *%*%*%*%*%*%*%*%*
\begin{figure}[tbp]
\includegraphics[width=8.5cm]{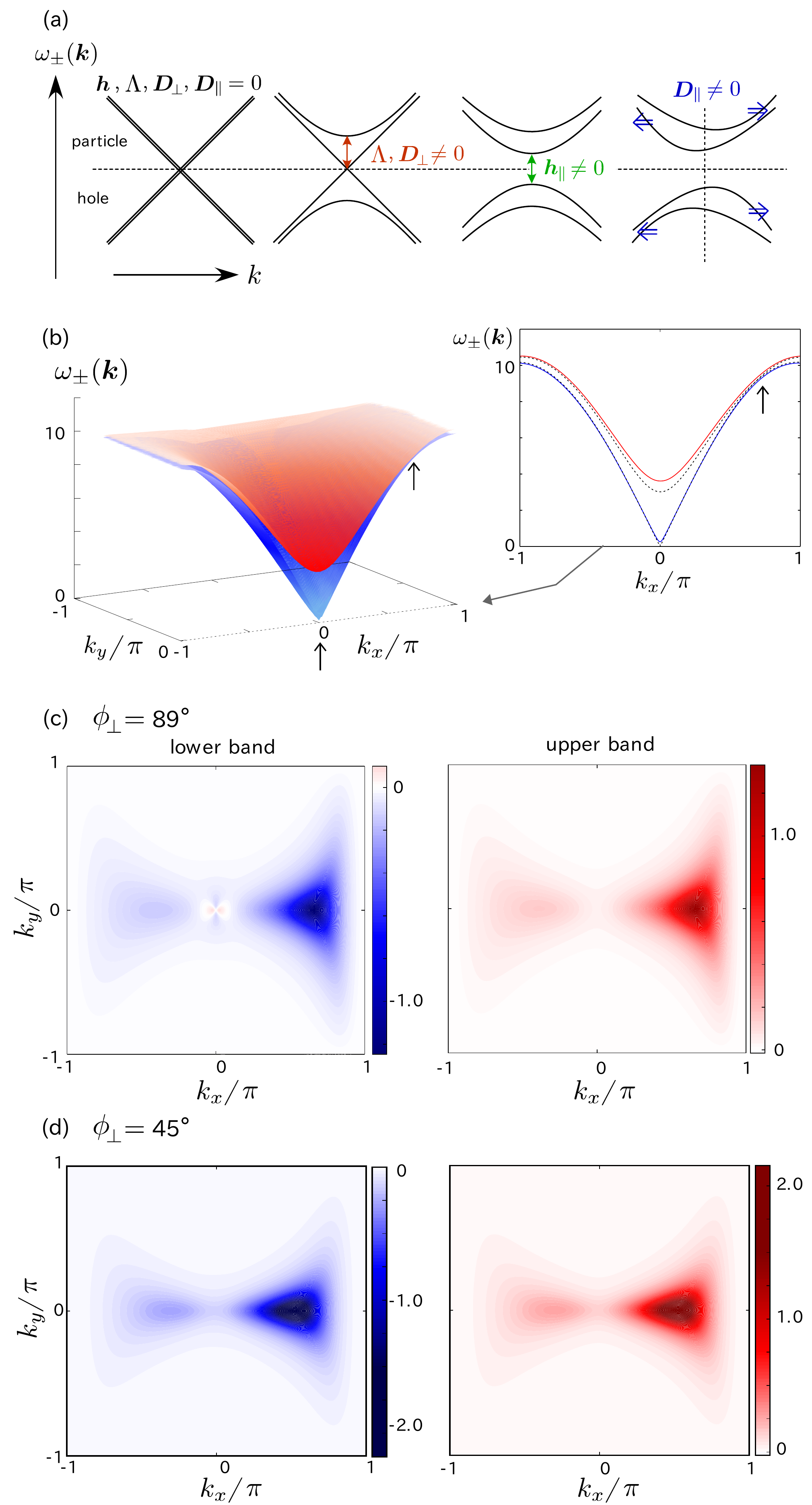}
\caption{(a) Schematic illustration of how the magnon bands are modified by the model parameters. 
(b) Magnon bands $\omega_\pm(\bm k)$ of $k_y \leq0$ at $D_\parallel=D_\perp=0.2$, $\Lambda=0.05$, 
$\phi_{\parallel}=0$, $\phi_{\perp}=89^{\circ}$. 
The right panel shows the cross sections of the magnon bands along the $k_{y}=0$ line, 
where the broken and solid lines are $\bm h_\perp=\bm{0}$ and $\ne \bm{0}$ cases, respectively. 
When $\theta_{\perp}=0$ ($\bm h_\perp=0$), there is a band touching point along the $k_{x}$-axis. 
The finite $\theta_{\perp}$ opens a gap between upper and lower bands.
(c)-(d) Density plot of the Berry curvature of the lower and upper bands, 
where we take (c)$\phi_{\perp}=89^{\circ}$ and (d) 45$^\circ$. }
\label{f3}
\end{figure}
%*%*%*%*%*%*%*%*%*%*%*%*%*%*%*%*%*%*
%*%*%*%*%*%*%*%*%*%*%*%*%*%*%*%*%*%*
%*%*%*%*%*%*%*%*%*%*%*%*%*%*%*%*%*%*
%*%*%*%*%*%*% fig5 *%*%*%*%*%*%*%*%*
\begin{figure}[tbp]
\includegraphics[width=8.5cm]{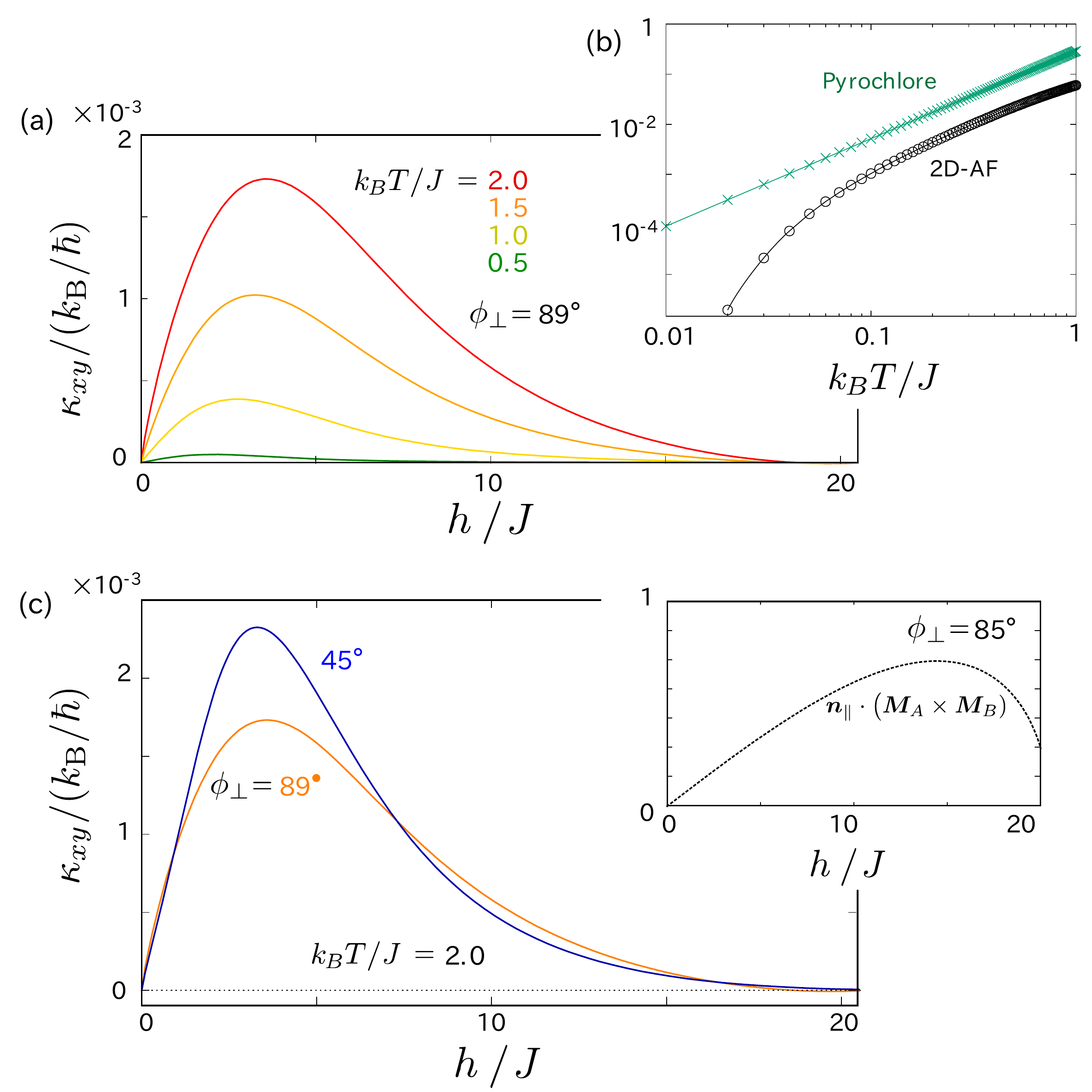}
\caption{In-plane thermal Hall conductivity $\kappa_{xy}/(k_{\mathrm{B}}/\hbar)$. 
(a) Magnetic field $h=|\bm{h}|$ dependence at several temperatures when $\phi_{\perp}=89^{\circ}$, 
(b) the temperature dependence at $\phi_{\perp}=89^{\circ}$, 
and (c) the comparison for two different field angles $\phi_{\perp}=45^{\circ}$ and $89^{\circ}$. 
(b) The comparison with the case of pyrochlore magnets in Ref.[\onlinecite{onose12}] 
that follows $\propto T^{7/2}$, which we calculated for $J=1.0$, $S=1/2$, $D=0.32$, and $H=+0$ using their formula. 
The DM interactions defined in Fig. 2(a) have the Dresselhaus-type geometry. 
The Rashba types of DM interactions are obtained by converting the directions of the uniform DM vectors 
along the $\bm \delta_2$-directions 
upside down, in which case $\kappa_{xy}$ becomes negative with the same amplitude. }
\label{f4}
\end{figure}
%*%*%*%*%*%*%*%*%*%*%*%*%*%*%*%*%*%*
%*%*%*%*%*%*%*%*%*%*%*%*%*%*%*%*%*%*
%
%*%*%*%*%*%*%*%*%*%*%*%*%*%*%*%*%*%*%*%*%*%*%*%*
\vspace{5mm}
\\
\section{Results}
%*%*%*%*%*%*%*%*%*%*%*%*%*%*%*%*%*%*%*%*%*%*%*%*
%*%*%*%*%*%*%*%*%*%*%*%*%*%*%*%*%*%*%*%*%*%*%*%*
\subsection{Berry curvature} 
\label{sec:berry}
%*%*%*%*%*%*%*%*%*%*%*%*%*%*%*%*%*%*
Let us demonstrate the emergent Berry curvature and the thermal Hall conductivity 
in our SIS broken square lattice antiferromagnet. 
The Berry curvature is activated
when the spins cant in the out-of-plane direction due to $\bm{h}_{\perp}\neq\bm{0}$,
following the mechanism we discussed in \S.\ref{sec:symm}.
\par
We first show in Figs.~\ref{f3}(a) and \ref{f3}(b) a schematic illustration of the magnon bands and 
the typical magnon dispersion of Eq.(\ref{ham}), respectively. 
In the Heisenberg antiferromagnet with $D_{\perp}=D_{\parallel}=0$, $\Lambda=0$, and $\bm{h}=\bm{0}$, 
a well-known doubly degenerate dispersion appears which is linear in $k$ around the $\Gamma$-point. 
When $\Lambda >0$, the degeneracy is lifted and only the lower band remains zero at the $\Gamma$-point.
By the introduction of $\bm D_\perp$ (or either $\bm h_\parallel$), the spins are canted in-plane 
($\theta_\parallel \ne 0)$, and due to $\bm D_\parallel\ne 0$ the nonreciprocity appears \cite{kawano18}. 
At $h_\perp =0$, there is an accidental band touching point\cite{asano-hotta} along the $k_x$-axis ($k_y=0$) indicated by arrows in Fig.\ref{f3}(b), 
besides the trivial degeneracy with the hole bands on the bottom ($\Gamma$-point).
The former degeneracy is lifted by $\bm{h}_{\perp}\neq\bm{0}$,
and the latter by $\bm{h}_{\parallel}\neq\bm{0}$.
\par
We show in Fig.~\ref{f3}(c) the density plot of the Berry curvature for $\phi_{\parallel}=0$ and $\phi_{\perp}=89^{\circ}$ 
(the magnetic field almost perpendicular to the plane).
One can see the enhancement of $\Omega_{xy}^{(\pm)}(\bm{k})$ 
at finite $k_x$ off the $\Gamma$-point. 
This enhancement takes place in the vicinity of the $\bm{k}$-point 
where the band touching is observed at $\bm{h}_{\perp}=\bm{0}$.
Then the two bands, which basically have different symmetry in the contributions from A- and B-sublattices, 
mix and generate a large $\Omega_{xy}^{(\pm)}(\bm{k})$. 
When the gap is small, the degree of the band mixing and the value of $\Omega_{xy}^{(\pm)}(\bm{k})$
are large, which are gradually suppressed by the increase of $\bm{h}_{\perp}$. 
In contrast, the particle-hole contact at the $\Gamma$-point 
does not contribute much to $\Omega_{xy}^{(\pm)}(\bm{k})$ when $\bm{h}_{\perp}\ne 0$,
since these two bands have the same symmetry 
in terms of A- and B-sublattice degrees of freedom.
\par
We also show in  Fig.~\ref{f3}(d) the Berry curvature at $\phi_{\perp}=45^{\circ}$. 
The in-plane element, $\bm{h}_\parallel$, first works to shift the (nearly) band-touching point in the $-k_x$-direction. 
Therefore at this field angle, the region where $\Omega_{xy}^{(\pm)}(\bm{k})$ takes the maximum shifts to lower energy levels, 
and its amplitude is enhanced (since $h_\perp=|\bm{h}_{\perp}|=h/\sqrt{2}<h$). 
A further increase of $h_\parallel$ shifts the band-touching point from the center toward the other edge of the Brillouin zone, 
and suppresses $\Omega_{xy}^{(\pm)}(\bm{k})$.
The details of how the Berry curvature develops when the magnetic field is varied 
is shown in Appendix \ref{app4}. 
%
%*%*%*%*%*%*%*%*%*%*%*%*%*%*%*%*%*%*
%*%*%*%*%*%*%*%*%*%*%*%*%*%*%*%*%*%*
\begin{figure*}[tb]
\includegraphics[width=17cm]{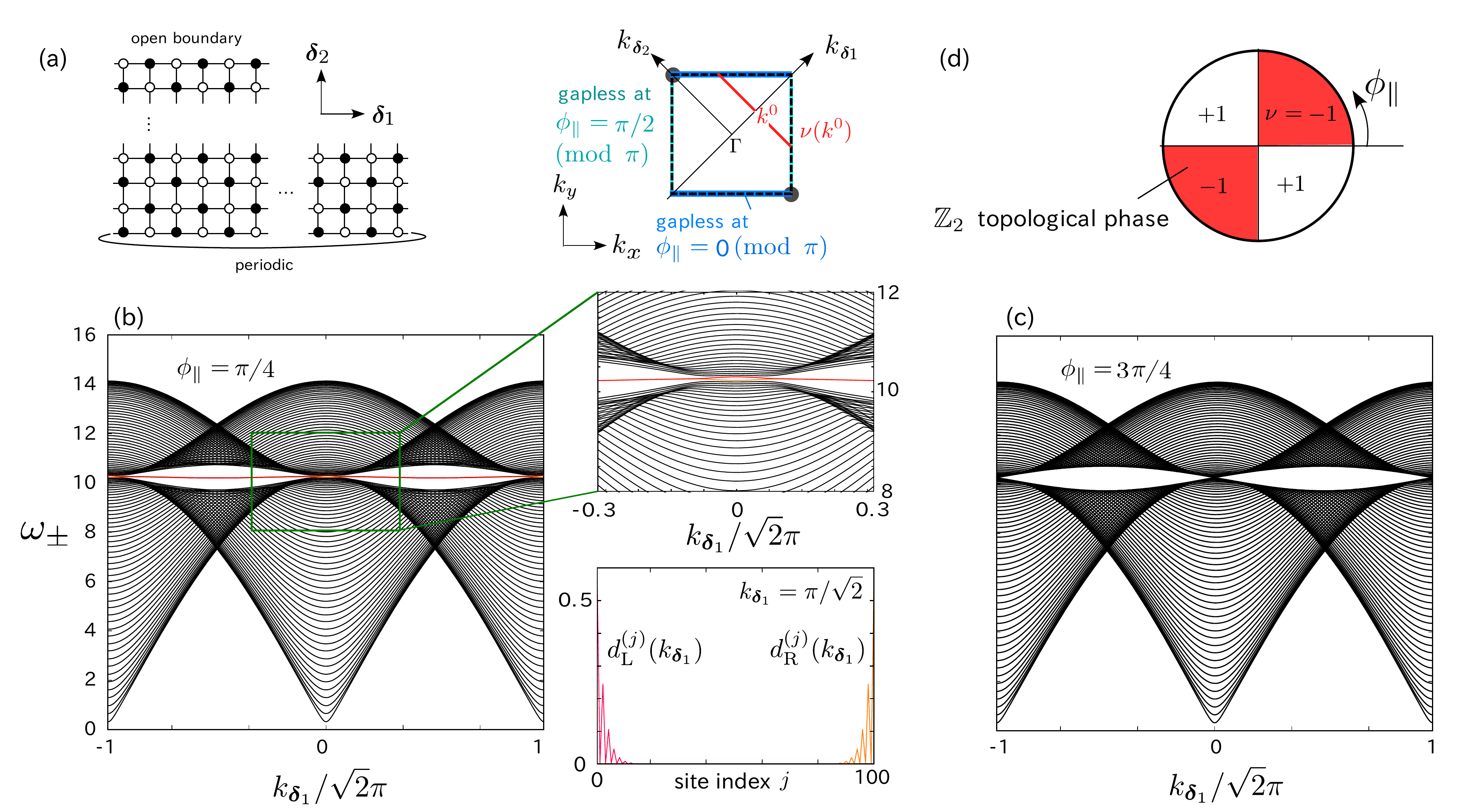}
\caption{(a) Lattice with periodic boundary in the $\bm \delta_1$-direction and open boundary at $\bm \delta_2=0$ and $2L$. 
The topological invariant is defined independently from the calculation of the edge modes 
along $k_{\bm{\delta}_2}$ line for each fixed $k_{\bm{\delta}_1}$ in the Brillouin one, 
avoiding the gapless point at $k_{\bm{\delta}_1}=0,\pm\sqrt{2}\pi$. 
(b), (c) The energy dispersion of magnons, $\omega_\pm(k_{\bm{\delta}_1})$, consisting of $2L$ lines, 
calculated for the lattice in panel (a). 
The red/orange solid lines present only in (b) are the topologically protected edge modes. 
We take $L=50$, $D_\perp = 0.2$, $D_\parallel=0.2$, $\Lambda= 0.05$,  $h=14.0$, with 
$\phi_\perp=89^{\circ}$ 
and (b)$\phi_\parallel=\pi/4$ and (c)$\phi_\parallel=3\pi/4$. 
The levels near the modes are magnified in the middle panel, and in the lower middle panel, 
the local density of states of the two edge modes, 
$d^{(j)}_{L/R}(k_{\bm \delta_1}=\pi/\sqrt{2})$ along the 
$\bm \delta_2$-direction in real space are shown.
(d) Phase diagram as a function of $\phi_\parallel$ 
classifying the $\nu=\pm 1$ regions, corresponding to the topological and trivial phases.
}
\label{f5}
\end{figure*}
%*%*%*%*%*%*%*%*%*%*%*%*%*%*%*%*%*%*
%*%*%*%*%*%*%*%*%*%*%*%*%*%*%*%*%*%*
\subsection{Thermal Hall effect}
\label{sec:thermal}
%*%*%*%*%*%*%*%*%*%*%*%*%*%*%*%*%*%*
The thermal Hall conductivity Eq.(\ref{eq:Thermal-Hall}) is calculated based on 
the above mentioned band structures. 
Figure \ref{f4} shows $\kappa_{xy}$
as a function of magnetic field $h$ for several different temperatures. 
The enhancement of $\kappa_{xy}$ by $k_BT$ is due to the thermal excitation following the bosonic distribution function. 
The comparison with the thermal Hall conductivity 
obtained for the previously reported pyrochlore ferromagnet \cite{onose12} is shown in Fig.~\ref{f4}(b); 
the pyrochlore ferromagnet has only a small magnetic field dependence, 
and a clear $T^{7/2}$-dependence is observed. 
Notice that the $k_BT$ is scaled by $J$, which depends much on the material parameters (also $D$ as well), 
 so that the direct comparison of the absolute values of $\kappa_{xy}$
between the two systems does not make sense. 
The distinct feature of our antiferromagnet is the rapid growth of $\kappa_{xy}$, 
beyond the simple power of $T$ at temperatures lower than $k_BT \lesssim 0.1J$, 
which is actually the target temperature range of measurements in laboratories. 
This should come from the nearly gapless excitation of the lower magnon branch, 
where the tail of the peak of $\Omega_{xy}^{(\pm)}(\bm{k})$ 
extending from the band touching point at higher energy has a non-negligible contribution. 
The gapped pyrochlore ferromagnet does not have such low energy modes. 
In fact, as the temperature increases, $\kappa_{xy}$ of our antiferromagnets extrapolates to 
$T^{7/2}$-behavior of the pyrochlore ferromagnets, 
where the contribution from the Berry curvature at the higher magnon bands dominates. 
\par
Figure~\ref{f4}(c) shows the $h$-dependence of $\kappa_{xy}$ for another field angle 45$^\circ$. 
At $h=0$, we have exactly $\kappa_{xy}=0$ since $\theta_\perp=0$. 
When a finite $h$ (with $\phi_\perp >0$) 
is introduced, the spins start to cant and the solid angle shown in Fig.\ref{f2}(b) 
increases and so as $\Omega_{xy}^{(\pm)}(\bm{k})$, which explains the rapid growth of $\kappa_{xy}$. 
The functional form of the solid angle at small $h$ given in the inset actually resembles that of $\kappa_{xy}$. 
As we discussed in Figs.\ref{f4}(c) and \ref{f4}(d), 
rotating the angle from 89$^\circ$ to 45$^\circ$ shifts the point with 
maximum $\Omega_{xy}^{(\pm)}(\bm{k})$ to lower energy levels, which then works to enhance $\kappa_{xy}$. 
The further increase of field or field angle then pushes this maximum point to higher energies 
(see Appendix D) 
and $\kappa_{xy}$ becomes gradually suppressed, even though the solid angle continues to increase 
up to $h/J \sim 15$. 
%
%*%*%*%*%*%*%*%*%*%*%*%*%*%*%*%*%*%*%*%*%*%*%*%*
\subsection{Edge mode in the $\mathbb{Z}_2$ topological phase} 
\label{sec:topo}
We finally show the existence of topological phase characterized by 
the $\mathbb{Z}_2$ topological invariant.
For simplicity, we set the field nearly perpendicular to the plane, $\phi_\perp=89^{\circ}$. 
The in-plane direction of the classical spins in the ground state is 
determined by a small $\bm{h}_{\parallel}$.
We show that $\phi_\parallel$ plays an important role.
Note that the magnitude of $\bm{h}_{\parallel}$ is not important as it simply serves to determine the in-plane angle of the magnetic moment.
\par 
The calculation on edge modes (see \S.\ref{sec:z2}) are based on the lattice of width $2L$ with open edges.
As shown in Fig.\ref{f5}(a), we take the periodic boundary in the $\bm{\delta}_{1}$-direction 
and obtain $2L$ sets of particle and hole bands defined along the $k_{\bm{\delta}_{1}}$-direction in momentum space. 
Figures~\ref{f5}(b) and \ref{f5}(c) display the particle bands, 
$\omega_\pm(k_{\bm{\delta}_{1}})\geq 0$
for $\phi_{\parallel}=\pi/4$ and $3\pi/4$, respectively. 
Although they look alike, only the former affords topologically protected edge modes (red and orange lines) inside the gap. 
The topologically protected edge modes disappear at $k_{\bm{\delta}_{1}}=0,\pm\sqrt{2}\pi$,
which corresponds to the corner of the Brillouin zone,
where the band gap closes.
The distributions of these edge states in real space 
described by the local density of states $d_{L/R}^{(j)}(k_{\bm{\delta}_{1}})$, 
are shown in the inset of Fig.\ref{f5}(b)(bottom panel in the middle) 
along the $\bm \delta_2$-direction. 
[For details of $d_{L/R}^{(j)}(k_{\bm{\delta}_{1}})$ see Appendix A.] 
One can see that the left and right edge states are localized over $\sim 5$ lattice spacings 
from the open edges. 
\par
We now examine the topological invariant to confirm the above findings.
When $\phi_{\parallel}= n\pi/2$ (with integer $n$), 
there is always a band degeneracy 
along either the $k_x=\pm\pi$ line ($n=$odd) 
or the $k_y=\pm\pi$ line ($n=$even) at the Brillouin zone boundary. 
Otherwise, the bands are gapped except at the corner of the Brillouin zone, 
and we can define a winding number as in Eq. (\ref{eq:wind}).
The winding number is obtained as,
\begin{equation}
W(k_{\bm{\delta}_{1}})=\pi M,\hspace{5mm} M \in \mathbb{Z} 
\end{equation}
and accordingly, the $\mathbb{Z}_{2}$ topological invariant becomes, 
\begin{equation}
\nu(k_{\bm{\delta}_{1}})=\left\{
\begin{array}{l}
-1 \hspace{10pt} (\phi_{\parallel}=(0,\frac{\pi}{2}),\;(\pi,\frac{3\pi}{2})) \\
\ \\
+1 \hspace{10pt} (\phi_{\parallel}=(\frac{\pi}{2},\pi),\;( \frac{3\pi}{2},2\pi))
\end{array}
\right.
\end{equation}
The phase diagram regarding $\phi_\parallel$ is shown in Fig.\ref{f5}(d); 
there is a topological phase transition at $\phi_\parallel=n \pi/2$ (with integer $n$), 
where the band degeneracy actually takes place, 
and the diagram is separated into the topological and trivial regions. 
Thus, regarding Figs.\ref{f5}(b) and \ref{f5}(c), 
only the former has the topologically protected edge modes. 
%
%*%*%*%*%*%*%*%*%*%*%*%*%*%*%*%*%*%*
%*%*%*%*%*%*% fig6 *%*%*%*%*%*%*%*%*
\begin{figure}[tbp]
\includegraphics[width=8.4cm]{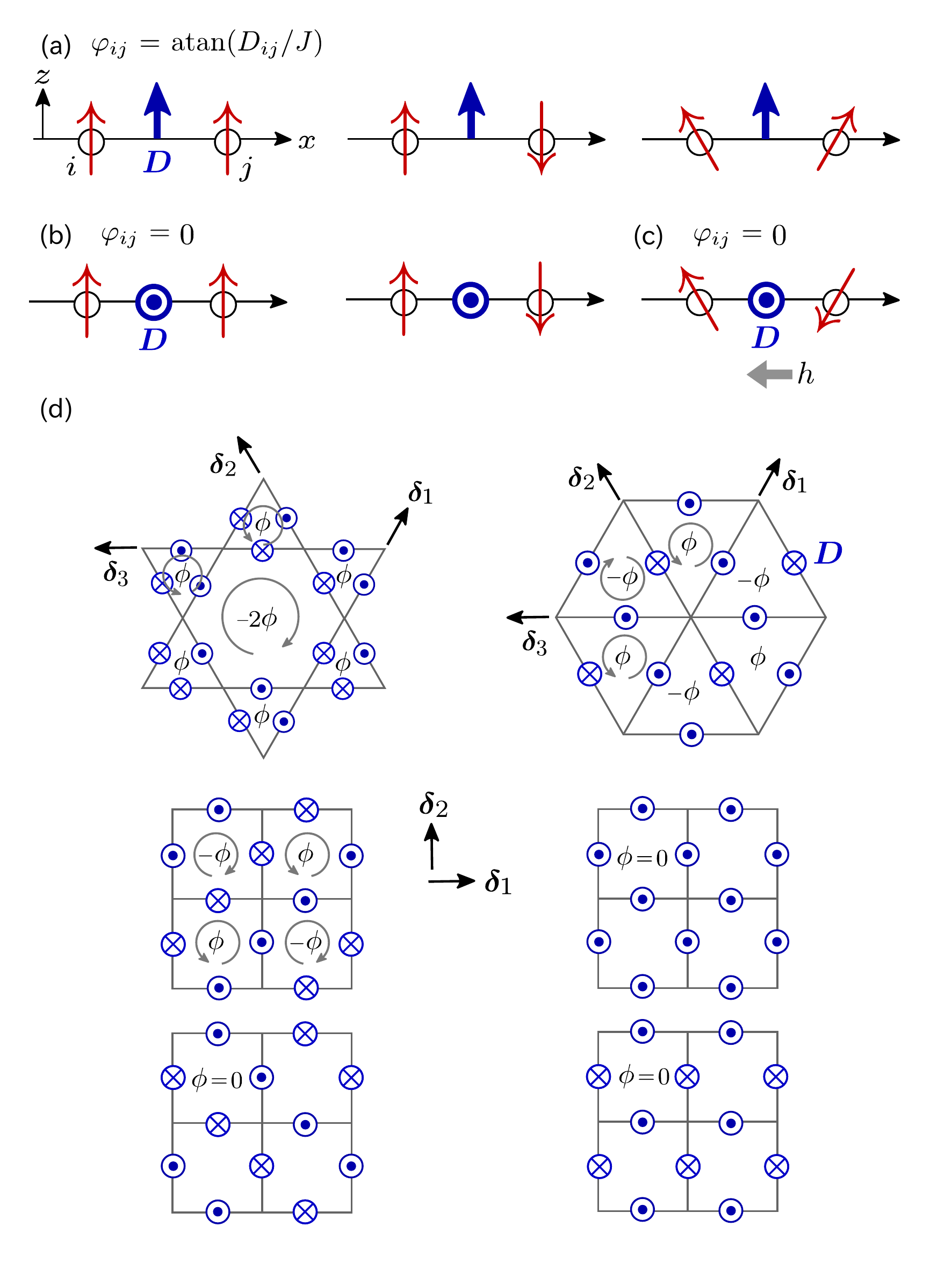}
\caption{(a)-(c) Schematic illustration of the coupling of DM interactions 
with collinear or nearly collinear magnetic structures. 
The representative cases where the magnetic moments are (a) parallel and 
(b) perpendicular to the DM interaction, 
where the former affords a local phase $\varphi_{ij}$, but the latter does not.
(c) When the antiferromagnetic moments cant within the plane perpendicular to $\bm D$, 
we find a finite Berry curvature although $\varphi=0$, which is explained in Fig.\ref{f2}. 
(d) The 2D lattices with DM interactions.
The flux inside the closed loop $\cal C$ is defined as, $\phi=\sum_{\cal C} \varphi_{ij}$. 
The net flux is zero in all cases. 
The thermal Hall effect is present only in the kagome lattice,
since it contains two inequivalent loops in the unit cell.
}
\label{f6}
\end{figure}
%*%*%*%*%*%*%*%*%*%*%*%*%*%*%*%*%*%*
%*%*%*%*%*%*%*%*%*%*%*%*%*%*%*%*%*%*
%*%*%*%*%*%*%*%*%*%*%*%*%*%*%*%*%*%*%*%*%*%*%*%*
%*%*%*%*%*%*%*%*%*%*%*%*%*%*%*%*%*%*%*%*%*%*%*%*
\section{discussion}
\label{sec:discuss}
%*%*%*%*%*%*%*%*%*%*%*%*%*%*%*%*%*%*%*%*%*%*%*%*
%*%*%*%*%*%*%*%*%*%*%*%*%*%*%*%*%*%*%*%*%*%*%*%*
We demonstrated that the thermal Hall conductivity and topologically protected edge modes appear 
due to the nonvanishing Berry curvature in a square lattice antiferromagnet, 
when the magnetic moments cant and point in the direction perpendicular to 
the DM vector as shown in Fig.\ref{f0}(b). 
We stressed in \S.\ref{sec:intro} that the conventional mechanism using a U(1) gauge field in Fig.\ref{f0}(a) does not apply to square lattice models. 
To further clarify this point, 
we classify the role of DM interactions by the way they align and by the lattice geometries 
based on the established idea given in Refs.[\onlinecite{onose12}] and [\onlinecite{katsura10}]. 
\par
Let us discuss the simplest part of the Hamiltonian including only the Heisenberg exchange and the DM interaction, 
\begin{eqnarray}
{\cal H}_{ij}&=&J\bm{S}_{i}\cdot\bm{S}_{j}+D_{ij}^{z}\bm{e}_{z}\cdot[\bm{S}_{i}\times\bm{S}_{j}]
 \nonumber \\
\hspace{20pt}&=& J_{\rm eff} ({\rm e}^{-i \varphi_{ij}}S_i^+ S_j^-+ {\rm e}^{i \varphi_{ij}}S_i^- S_j^+) 
+ JS_{i}^z S_{j}^z,
\rule{5mm}{0mm}
\end{eqnarray}
with $\varphi_{ij}={\rm atan}(D_{ij}^{z}/J)$ and $J_{\rm eff}=\sqrt{J^2+(D_{ij}^{z})^2}/2$.
The transfer integral of magnons, $J_{\rm eff}{\rm e}^{i \varphi_{ij}}$, 
has a finite phase factor when $D_{ij}^z\ne 0$. 
\par
Figures \ref{f6}(a)-\ref{f6}(c) show the relative relationships between the ordered magnetic moments and the DM vectors. 
When the DM vector has a finite element parallel to the ordered magnetic moments, 
the kinetics of magnons couples to $\bm D$ and gains a phase $\varphi_{ij}$. 
Here, we stress that when $\bm D$ is perpendicular to the moments [Figs.\ref{f6}(b) and \ref{f6}(c)], 
this mechanism does not hold
as we showed explicitly in Figs.\ref{f2}(d) and \ref{f2}(e). 
\par
Now, we consider 2D lattices shown in Fig.\ref{f6}(d). 
Since the sign of $\bm D_{ij}$ changes by converting $i$ and $j$, 
we define the direction $i\rightarrow j$ 
as $\bm \delta_\gamma$, $\gamma=1,2,3$, in the figure to fix the direction in the figure. 
Here, we prepare several types of lattices with  
the staggered and uniform DM interactions [see Figs.~\ref{f1}(b) and \ref{f1}(c)]: 
a kagome and a triangular lattice for the former and the latter, respectively, 
and a square lattice for both. 
Let us take a closed loop, ${\cal C}$, along the bonds in each of these lattices. 
When $\varphi_{ij}\ne 0$, a fictitious magnetic flux, 
$\phi=\sum_{\mathcal{C}}\varphi_{ij}$, is generated, 
where we take the counterclockwise direction in the summation. 
The net flux is zero in all cases but the thermal Hall conductivity is finite in the kagome lattice 
and zero in the square and triangular lattices. 
\par
The reason is explained as follows.
In the square and triangular lattices, 
there are geometrically equivalent loops in the unit cell, 
and there exists a symmetry operation that does not change the flux pattern 
but converts the sign of the thermal Hall conductivity. 
This straightforwardly implies that the thermal Hall conductivity is zero. 
Note that this symmetry operation does not change the effective Hamiltonian of free bosons in the presence of fluxes.
\par
This argument is illustrated more precisely as follows.
Let us consider a general $2n\times2n$ BdG Hamiltonian, $H_{\mathrm{BdG}}(\bm{k})$, 
of magnons where $n$ denotes the number of degrees of freedom in a unit cell. 
Suppose that there is an operator represented by the $2n\times2n$ unitary matrix, 
$\cal O$,  
describing the above mentioned symmetry operation in the system, 
which satisfies, 
\begin{equation}
\mathcal{O}KH_{\mathrm{BdG}}(\bm{k})K^{-1}\mathcal{O}^{\dagger}=H_{\mathrm{BdG}}(\bm{k}),
\label{eq:sym-bdg}
\end{equation}
together with $[\mathcal{O},\tilde{\Sigma}^{z}]=0$ 
and $\tilde{\Sigma}^{z}=\tau^{z}\otimes1_{n\times n}$. 
For example, $\mathcal{O}K$ can be the exchange of the sublattices and the conversion of the flux associated with that operation. 
If Eq.(\ref{eq:sym-bdg}) is fulfilled and 
if the energy bands are not degenerate at $\bm k$, 
one can take the eigenvector, $\bm{t}_{\mu}(\bm{k})$ ($\mu=1,2,\cdots,2n$), 
of $H_{\mathrm{BdG}}(\bm{k})$ in the following form, 
\begin{equation}
\bm{t}_{\mu}(\bm{k})=\mathrm{e}^{i\varphi_{\mu}(\bm{k})}\mathcal{O}\bm{t}_{\mu}^{*}(\bm{k}),
\label{eq:symt}
\end{equation}
where $\varphi_{\mu}(\bm{k})\in\mathbb{R}$ is some proper choice of a phase factor.
From Eq.(\ref{eq:symt}) and from the normalization condition, 
$\bm{t}_{\mu}^{\dagger}(\bm{k})\tilde{\Sigma}^{z}\bm{t}_{\mu}(\bm{k})
=(\tilde{\Sigma}^{z})_{\mu,\mu}$, it is straightforward to see that 
the Berry curvature of the $\mu$-th band satisfies
\begin{align}
\Omega_{xy}^{(\mu)}(\bm{k})&=
-2(\Sigma^{z})_{\mu,\mu}\mathrm{Im}
\left[
\frac{\partial\bm{t}_{\mu}^{\dagger}(\bm{k})}{\partial k_{x}}
\tilde{\Sigma}^{z}\frac{\partial\bm{t}_{\mu}(\bm{k})}{\partial k_{y}}
\right]
\nonumber \\
&=
-2(\Sigma^{z})_{\mu,\mu}\mathrm{Im}
\left[
\frac{\partial\bm{t}_{\mu}^{T}(\bm{k})}{\partial k_{x}}\mathcal{O}^{\dagger}\tilde{\Sigma}^{z}\mathcal{O}
\frac{\partial\bm{t}_{\mu}^{*}(\bm{k})}{\partial k_{y}}
+(\mathrm{real})
\right]
\nonumber \\
&=-\Omega_{xy}^{(\mu)}(\bm{k}).
\end{align}
In this way, we find $\Omega_{xy}^{(\mu)}(\bm{k})=0$ 
when there exists an operator $\mathcal{O}$ that satisfies Eq.(\ref{eq:sym-bdg}).
The actual example is discussed previously in Ref.[\onlinecite{onose12}]. 
From Eqs.(\ref{eq:sis}) and (\ref{eq:ptrs}), 
our antiferromagnet {\it without} $\bm D_{\parallel}$ satisfies Eq.(\ref{eq:sym-bdg})
with $\mathcal{O}=(\tau^{0}\otimes\sigma^{x})$,
that makes $\Omega_{xy}^{(\mu)}(\bm{k})=0$.
\par
Now, we go back to the lattices shown in Fig.\ref{f6}(d).
In the square and triangular lattices with finite fluxes, 
a $\pi$ rotation around the $\bm{\delta}_{1}$ axis 
connects different loops of the same size and shape and serves as $\mathcal{O}$. 
For the three cases of a square lattice in the lower part of the figure, the Berry curvature is already zero,
whereas in the kagome lattice, 
there is no such symmetry operation $\mathcal{O}$ 
that connects inequivalent loops in the unit cell, 
and the Berry curvature remains finite\cite{katsura10,onose12}.
\par
The above discussion was previously applied to centrosymmetric ferromagnets with SIS. 
The SIS broken systems with uniform DM interaction were thus overlooked. 
This shall be because, if the SIS is broken and the DM interaction aligns uniformly [Fig.\ref{f2}(c)] 
the flux itself is absent as we see in the lower right panels of Fig.\ref{f6}(d). 

%*%*%*%*%*%*%*%*%*%*%*%*%*%*%*%*%*%*%*%*%*%*%*%*
%*%*%*%*%*%*%*%*%*%*%*%*%*%*%*%*%*%*%*%*%*%*%*%*%*%*%*%*%*%*%*%*%*%*%*%*%*%*
\section{Summary}
We showed theoretically that the 2D inversion-symmetry broken square lattice antiferromagnet 
hosts the thermal Hall effect and the topologically protected edge modes. 
These two properties are based on the finite Berry curvature of magnon bands. 
The conventionally established way to generate a finite Berry curvature was to 
generate a U(1) gauge field or a fictitious flux discussed in the previous section.
The $\bm D$-vector {\it parallel} to the moments couples to the magnons as a vector potential and generates a local phase in their hoppings. 
By the proper choice of the geometry of lattices, which are the 
corner shared kagome and pyrochlore, 
one could avoid the cancellation of the effect of the U(1) gauge field. 
However, within that framework, our square lattice geometry can never afford a nonvanishing Berry curvature, 
since there exists a symmetry operation that cancels out the effect of U(1) gauge field together with the TRS.
\par
In our antiferromagnet, a different mechanism works; 
there are two species of magnons for two sublattices, and they serve as a pseudo-spin degrees of freedom. 
When a magnon hops from the A-sublattice to the neighboring B-sublattice, 
it accompanies a pseudo-spin flip, represented by the Pauli matrix, $i \sigma_y$. 
When the SIS is broken, the hoppings of magnons to the left and right neighbors become antisymmetric and couple to the DM vector via opposite sign as $-i\sigma_y$ and $+i\sigma_y$, respectively. 
It resembles the antisymmetric hopping of the Rashba electrons accompanying a spin flip in the SU(2) gauge field, where an anomalous Hall effect occurs from the emergent Berry curvature of electron bands. 
Therefore, our uniform DM vector can be regarded as a pseudo-spin orbit coupling of magnons. 
This picture is rather simplified but makes it straightforward to understand why our system generates a nonvanishing Berry curvature. 
We showed in more detail that to activate the couplings with the DM vector, 
one needs to control the antiferromagnetic moments to point in the direction 
{\it perpendicular} to the $\bm D$-vectors, and to make it noncollinear 
by the magnetic field. 
\par
The touching point (with a tiny gap) of two magnon bands 
appears at the middle of the energy bands, which is the point where the Berry curvature is enhanced. 
Since the antiferromagnetic magnon bands have low energy branches near the zero energy level, 
the tail of the Berry curvature extending from this high energy point allows for the rapid increase of the thermal Hall conductivity beyond the powers of $T$, 
in sharp contrast to the $T^{7/2}$-dependence in the pyrochlore ferromagnets. 
We also demonstrated that the topological phase is feasible in this system\cite{footnote}, 
and found that the topological phase transition characterized by the winding number is controlled by the rotation of the in-plane field angle.

\begin{acknowledgments}
We acknowledge discussions with Y. Onose and Y. Iguchi and Y. Nii and K. Penc. This work is supported by JSPS KAKENHI 
Grants Numbers No. JP17K05533, No. JP17K05497, No. JP17H02916, and No.JP18H01173. 
\end{acknowledgments}

%*%*%*%*%*%*%*%*%*%*%*%*%*%*%*%*%*%*%*%*
%*%*%*%*%*%*%*%*%*%*%*%*%*%*%*%*%*%*%*%*

\appendix

%*%*%*%*%*%*%*%*%*%*%*%*%*%*%*%*%*%*%*%*%*%*%*%*%*%*%*%*%*%*%*%*%*
\begin{widetext}

\section{Spin texture in the ground state}
\label{app1}
Our theory is built on the classical ground state with two-sublattice antiferromagnetic ordering.
To verify the stability of this ground state,
we estimate the critical value of $D_{\parallel}$,
at which there exists a transition from our canted antiferromagnetic order to an incommensurate magnetic order.
We first numerically examine Eq.(\ref{ham}) and obtain the critical value of $D_{\parallel}$. 
Figure~\ref{fapp1}(a) shows the variation of lower magnon bands 
for different choices of $D_\parallel$, 
where the other parameters are fixed to the ones we used in the main text, 
$J=1.0$, $D_{\perp}=0.2$, $\Lambda=0.05$, and $h_\parallel=1.0$. 
With increasing $D_\parallel$ the bottom of the band at nonzero $\bm k$ goes down and 
reaches the zero energy level at the critical value, $D_\parallel \sim 0.363$. 
In Fig.\ref{fapp1}(b) the minimum of 
$\omega_-$ is plotted as a function of $D_\parallel$ for several choices of $h_\parallel$. 
The critical values of $D_\parallel$ are the points where these lines fall down to the zero energy level. 
These results indicate that the system takes the stable antiferromagnetic ground state for a wide range of parameters including the ones we chose. 
We also see that both $D_\perp$ and $h_\parallel$ stabilizes the canted antiferromagnetic ground state 
against $D_\parallel$. 
\par
One can further analytically obtain the instability point for simpler cases. 
Let us consider a 1D antiferromagnet with a uniform DM interaction and an easy-plane magnetic anisotropy,
whose Hamiltonian is given as,
\begin{equation}
\mathcal{H}=
J\sum_{j=1}^{N}\bm{S}_{j}\cdot\bm{S}_{j+1}+\sum_{j=1}^{N}\bm{D}_{\parallel}\cdot[\bm{S}_{j}\times\bm{S}_{j+1}]+\Lambda\sum_{j=1}^{N}(S_{j}^{z})^{2},
\label{eq:1dham}
\end{equation}
where $\bm{D}_{\parallel}=-D_{\parallel}\bm{e}_{y}$, 
and we take the periodic boundary condition.
The following discussion can be applied to our 2D antiferromagnet,
if we regard the $\bm{\delta}_{1}$- or $\bm{\delta}_{2}$-direction as $x$-direction in Fig.\ref{fapp1}(c).
\par
In the following, we neglect the staggered DM interaction, $D_{\perp}$, and the magnetic field, $\bm{h}$, 
for simplicity. 
Without the loss of generality,
one can assume that the direction of the ordered moments varies slowly in space,
and describe the classical moments as $\bm{S}_{i}=S\bm{n}_{\mathrm{A}}(x_{i})$ for 
the A-sublattice and $\bm{S}_{j}=S\bm{n}_{\mathrm{B}}(x_{j})$ for the B-sublattice,
where $\bm{n}_{\mathrm{A}/\mathrm{B}}(x)$ is the continuous field and satisfies $(\bm{n}_{\mathrm{A}/\mathrm{B}}(x))^{2}=1$.
Then the Hamiltonian is written as,
\begin{equation}
\mathcal{H}\simeq JS^{2}a\int_{0}^{L}dx\ 
\left[
\frac{1}{a^{2}}\bm{n}_{\mathrm{A}}\cdot\bm{n}_{\mathrm{B}}-\frac{1}{2}\partial_{x}\bm{n}_{\mathrm{A}}\cdot\partial_{x}\bm{n}_{\mathrm{B}}
-\frac{D_{\parallel}}{Ja}[\bm{n}_{\mathrm{A}}\times\partial_{x}\bm{n}_{\mathrm{B}}]^{y}
+\frac{\Lambda}{2Ja^{2}}\left\{(n_{\mathrm{A}}^{z})^{2}+(n_{\mathrm{B}}^{z})^{2}\right\}
\right],
\end{equation}
where $L=Na$ and $a$ is a lattice constant (see Fig.\ref{fapp1}(c)).
%*%*%*%*%*%*%*%*%*%*%*%*%*%*%*%*%*%*
%*%*%*%*%*%*%*%*%*%*%*%*%*%*%*%*%*%*
\begin{figure*}[tbp]
\includegraphics[width=18cm]{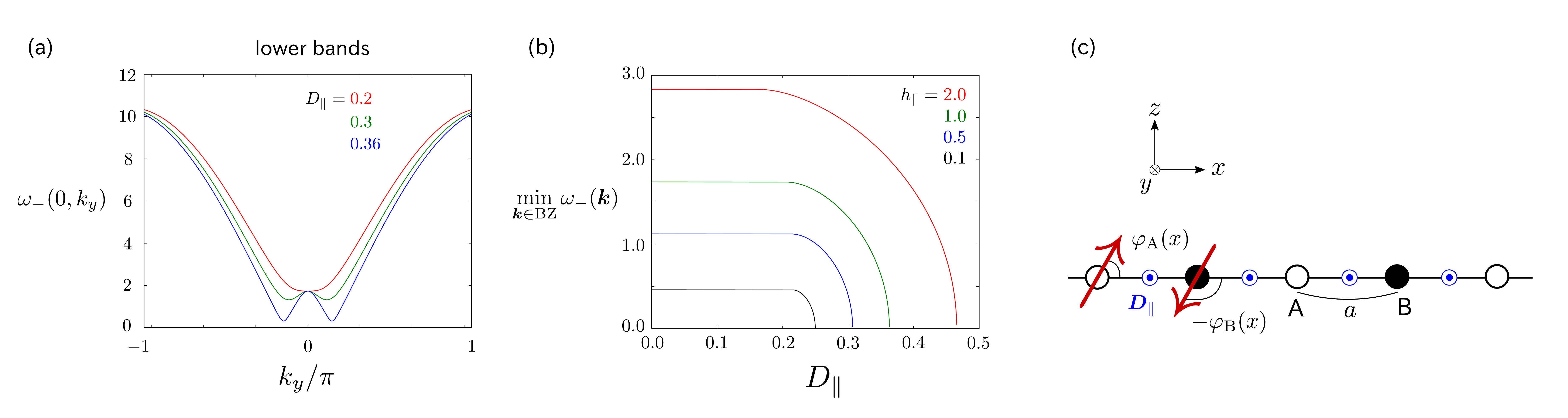}
\caption{ 
(a) Lower magnon bands of Eq.(\ref{ham}) for several different values of $D_{\parallel}$, 
where we take $J=1.0$, $D_{\perp}=0.2$, $\Lambda=0.05$, $h=1.0$, and $\phi_{\parallel}=\phi_{\perp}=0$. 
The instability of the canted two sublattice antiferromagnet to 
the incommensurate state occurs when the bottom of the bands fulfill 
$\omega_-(\bm k \ne 0)=0$, at $D_{\parallel}\simeq0.363$. 
(b) The  minimum energy of the magnon bands as a function of 
uniform DM interaction $D_{\parallel}$. The instability takes place when these lines 
fall down to zero. 
For all $\bm{h}$ with $D_{\perp}=D_{\parallel}=0.2$ 
and $\Lambda=0.05$, the two sublattice antiferromagnetic order is the stable ground state. 
(c) 1D antiferromagnetic chain with the uniform DM interaction and the easy-plane magnetic anisotropy.
}
\label{fapp1}
\end{figure*}
%*%*%*%*%*%*%*%*%*%*%*%*%*%*%*%*%*%*
For the collinear antiferromagnetic order within the easy-plane ($xy$-plane)
with uniformly distributed $\bm{n}_{\mathrm{A}}=-\bm{n}_{\mathrm{B}}$,
the energy is given by
\begin{equation}
E_{\mathrm{col}}=-JNS^{2}.
\end{equation}
Now we consider the case where the spins are pointing in the $xz$-plane.
The following calculation is similar to the one for the chiral soliton lattice\cite{kishine}.
The continuous field, $\bm{n}_{\mathrm{A}/\mathrm{B}}(x)$, can be written as
\begin{align}
\bm{n}_{\mathrm{A}}(x)&=
(\mathrm{cos}\varphi_{\mathrm{A}}(x),0,\mathrm{sin}\varphi_{\mathrm{A}}(x)),\\
\bm{n}_{\mathrm{B}}(x)&=
(\mathrm{cos}\varphi_{\mathrm{B}}(x),0,\mathrm{sin}\varphi_{\mathrm{B}}(x)),
\end{align}
and the Hamiltonian can be written using $\varphi_{\pm}(x)=\varphi_{\mathrm{A}}(x)\pm\varphi_{\mathrm{B}}(x)$ as
\begin{equation}
\mathcal{H}\simeq E_{\mathrm{col}}+JS^{2}a\int_{0}^{L}dx\ 
\left[
\left\{
-\frac{1}{8}(\partial_{x}\varphi_{+})^{2}+\frac{1}{8}(\partial_{x}\varphi_{-})^{2}+\frac{D_{\parallel}}{2Ja}(\partial_{x}\varphi_{+}-\partial_{x}\varphi_{-})
\right\}\mathrm{cos}\varphi_{-}
+\frac{\Lambda}{2Ja^{2}}(1-\mathrm{cos}\varphi_{+}\mathrm{cos}\varphi_{-})
\right].
\end{equation}
For the field $\varphi_{\pm}(x)$ representing a stationary point,
$\delta\mathcal{H}/\delta\varphi_{\pm}(x)=0$ is satisfied.
This gives, 
\begin{align}
\left(
\frac{1}{4}\partial_{x}^{2}\varphi_{+}+\frac{\Lambda}{2Ja^{2}}\mathrm{sin}\varphi_{+}
\right)\mathrm{cos}\varphi_{-}
-
\left(
\frac{1}{4}\partial_{x}\varphi_{+}-\frac{D_{\parallel}}{2Ja}\
\right)(\partial_{x}\varphi_{-})\mathrm{sin}\varphi_{-}
&=0,
\\
\left(
\frac{1}{8}(\partial_{x}\varphi_{+})^{2}+\frac{1}{8}(\partial_{x}\varphi_{-})^{2}-\frac{D_{\parallel}}{2Ja}(\partial_{x}\varphi_{+})+\frac{\Lambda}{2Ja^{2}}\mathrm{cos}\varphi_{+}
\right)\mathrm{sin}\varphi_{-}
-\frac{1}{4}(\partial_{x}^{2}\varphi_{-})\mathrm{cos}\varphi_{-}
&=0.
\end{align}
Assuming that $\bm{n}_{\mathrm{A}}(x)$ and $\bm{n}_{\mathrm{B}}(x)$ are locally antiparallel for all given $x$, i.e. $\varphi_{-}(x)=\pi$,
the above condition can be written as,
\begin{equation}
\frac{d^{2}}{dx^{2}}\varphi_{+}(x)=-\frac{2\Lambda}{Ja^{2}}\mathrm{sin}\varphi_{+}(x),
\end{equation}
and one finds,
\begin{equation}
\varphi_{+}(x)=2\mathrm{am}\left(\sqrt{\frac{2\Lambda}{Ja^{2}}}\frac{x}{m},m\right),
\end{equation}
where $\mathrm{am}(x,m)$ is the Jacobi amplitude function,
and the value $m$ satisfies $0\leq m\leq1$.
For a sufficiently long chain length $L$,
that satisfies $(\varphi_{\mathrm{A}}(L)-\varphi_{\mathrm{A}}(0))/2\pi\in\mathbb{N}$,
the energy is given by
\begin{equation}
E_{\mathrm{chiral}}(m)=E_{\mathrm{col}}+\Lambda NS^{2}\left(1-\frac{1}{m^{2}}\right)+\frac{2\Lambda NS^{2}}{mK(m)}
\left(
\frac{E(m)}{m}-\frac{\pi}{2}\frac{D_{\parallel}}{J}\sqrt{\frac{J}{2\Lambda}}
\right),
\end{equation}
where $K(m)$ and $E(m)$ are the complete elliptic integral of the first kind and second kind, respectively.
The derivative of $E_{\mathrm{chiral}}(m)$ with respect to $m$ is calculated as,
\begin{equation}
\frac{dE_{\mathrm{chiral}}(m)}{dm}=
-\frac{2\Lambda NS^{2}}{m^{2}(1-m^{2})}\frac{E(m)}{K^{2}(m)}
\left[
\frac{E(m)}{m}-\frac{\pi}{2}\frac{D_{\parallel}}{J}\sqrt{\frac{J}{2\Lambda}}
\right].
\end{equation}
The solution of $dE_{\mathrm{chiral}}(m)/dm|_{m=m^{*}}=0$
satisfies the following condition,
\begin{equation}
\frac{E(m^{*})}{m^{*}}=\frac{\pi}{2}\frac{D_{\parallel}}{J}\sqrt{\frac{J}{2\Lambda}},
\label{eq:kstar}
\end{equation}
and the ground state energy is given by
\begin{equation}
E_{\mathrm{chiral}}(m^{*})=
E_{\mathrm{col}}+\Lambda NS^{2}\left(1-\frac{1}{(m^{*})^{2}}\right)
\leq E_{\mathrm{col}}.
\end{equation}
Thus, if there exists a solution of Eq.(\ref{eq:kstar}), 
the ground state is no longer in a collinear antiferromagnetic state. 
The condition to exclude such solution is to have $D_{\parallel}$ as small as,
\begin{equation}
\frac{D_{\parallel}}{J}\leq\frac{2}{\pi}\sqrt{\frac{2\Lambda}{J}},
\label{eq:condition-Dpara}
\end{equation}
since $E(m)/m\geq1$.
When $J=1.0$ and $\Lambda=0.05$, 
Eq.(\ref{eq:condition-Dpara}) can be rewritten as $D_{\parallel}\leq 2/\sqrt{10}\pi\simeq0.2013$. 
If we introduce $D_\perp \ne 0$ and $h_\parallel \ne 0$, the critical value of $D_\parallel$ will increase. 
Therefore, it gives the lower bound of the phase boundary. 
The value $D_{\parallel}=0.2$ we adopted still remains below this lower bound.

%*%*%*%*%*%*%*%*%*%*%*%*%*%*%*%*%*%*%*%*%*%*%*%*%*%*%*%*%*%*%*%*%*
\section{Details of the spin wave analysis}
\label{app2}

We show the details of the results of the spin wave analysis in \S.\ref{sec:sw} in the main text.
The $2\times2$ matrices $\Xi_{\bm{k}}$ and $\Delta_{\bm{k}}$ in Eq.(\ref{eq:hk}) are given by,
\begin{align}
\Xi_{\bm{k}}&=
\xi^{(0)}\sigma^{0}+\xi_{\bm{k}}^{(x)}\sigma^{x}+\xi_{\bm{k}}^{(y)}\sigma^{y},
\\
\Delta_{\bm{k}}&=
\Delta^{(0)}\sigma^{0}+\Delta_{\bm{k}}^{(x)}\sigma^{x}+\Delta_{\bm{k}}^{(y)}\sigma^{y},
\end{align}
where
\begin{align}
\xi^{(0)}&=
4JS(\mathrm{cos}2\theta_{\parallel}\mathrm{cos}^{2}\theta_{\perp}-\mathrm{sin}^{2}\theta_{\perp})+4D_{\perp}S\mathrm{sin}2\theta_{\parallel}\mathrm{cos}^{2}\theta_{\perp}+\Lambda S(1-3\mathrm{sin}^{2}\theta_{\perp})
\nonumber \\
&\hspace{20pt}+h(\mathrm{sin}\phi_{\perp}\mathrm{sin}\theta_{\perp}+\mathrm{cos}\phi_{\perp}\mathrm{sin}\theta_{\parallel}\mathrm{cos}\theta_{\perp}),
\\
\xi_{\bm{k}}^{(x)}&=
-4JS\gamma_{\bm{k}}(\mathrm{sin}^{2}\theta_{\parallel}-\mathrm{cos}^{2}\theta_{\parallel}\mathrm{sin}^{2}\theta_{\perp})+4D_{\perp}S\gamma_{\bm{k}}\mathrm{sin}\theta_{\parallel}\mathrm{cos}\theta_{\parallel}(1+\mathrm{sin}^{2}\theta_{\perp})-2D_{\parallel}Sg_{1,\bm{k}}\mathrm{sin}\theta_{\parallel}\mathrm{cos}\theta_{\perp},
\\
\xi_{\bm{k}}^{(y)}&=
-4JS\gamma_{\bm{k}}\mathrm{sin}2\theta_{\parallel}\mathrm{sin}\theta_{\perp}+4D_{\perp}S\gamma_{\bm{k}}\mathrm{cos}2\theta_{\parallel}\mathrm{sin}\theta_{\perp}-D_{\parallel}Sg_{1,\bm{k}}\mathrm{cos}\theta_{\parallel}\mathrm{sin}2\theta_{\perp},
\\
\Delta^{(0)}&=
-\Lambda S\mathrm{cos}^{2}\theta_{\perp},
\\
\Delta_{\bm{k}}^{(x)}&=
4JS\gamma_{\bm{k}}\mathrm{cos}^{2}\theta_{\parallel}\mathrm{cos}^{2}\theta_{\perp}+2D_{\perp}S\gamma_{\bm{k}}\mathrm{sin}2\theta_{\parallel}\mathrm{cos}^{2}\theta_{\perp},
\\
\Delta_{\bm{k}}^{(y)}&=D_{\parallel}Sg_{1,\bm{k}}\mathrm{cos}\theta_{\parallel}\mathrm{sin}2\theta_{\perp}-2iD_{\parallel}Sg_{2,\bm{k}}\mathrm{cos}\theta_{\parallel}\mathrm{cos}\theta_{\perp},
\\
\gamma_{\bm{k}}&=\frac{\mathrm{cos}\bm{k}\cdot\bm{\delta}_{1}+\mathrm{cos}\bm{k}\cdot\bm{\delta}_{2}}{2},
\\
g_{1,\bm{k}}&=\mathrm{cos}\phi_{\parallel}\frac{\mathrm{sin}\bm{k}\cdot\bm{\delta}_{1}-\mathrm{sin}\bm{k}\cdot\bm{\delta}_{2}}{\sqrt{2}}-\mathrm{sin}\phi_{\parallel}\frac{\mathrm{sin}\bm{k}\cdot\bm{\delta}_{1}+\mathrm{sin}\bm{k}\cdot\bm{\delta}_{2}}{\sqrt{2}}
\\
g_{2,\bm{k}}&=\mathrm{cos}\phi_{\parallel}\frac{\mathrm{sin}\bm{k}\cdot\bm{\delta}_{1}+\mathrm{sin}\bm{k}\cdot\bm{\delta}_{2}}{\sqrt{2}}+\mathrm{sin}\phi_{\parallel}\frac{\mathrm{sin}\bm{k}\cdot\bm{\delta}_{1}-\mathrm{sin}\bm{k}\cdot\bm{\delta}_{2}}{\sqrt{2}}.
\end{align}
%*%*%*%*%*%*%*%*  eiven values %*%*%*%*%*%*%*%*
The final analytical form of the magnon band is given as, 
\begin{equation}
\omega_{\pm}(\bm{k})=\frac{1}{2}\left(\sqrt{w_{\bm{k}}}\pm\sqrt{2P_{\bm{k}}+\frac{2Q_{\bm{k}}}{\sqrt{w_{\bm{k}}}}-w_{\bm{k}}}\right)
\label{eq:omega}
\end{equation}
%*%*%*%*%*%*%*%*%*%*%*%*%*%*%*%*%*%*%*%*%*%*%*%*
where $w_{\bm{k}}$, $P_{\bm{k}}$, $Q_{\bm{k}}$ are given explicitly as, 
\begin{equation}
w_{\bm{k}}=\eta_{\bm{k}}+\zeta_{\bm{k}}+\frac{2}{3}P_{\bm{k}}\in\mathbb{R},
\end{equation}
\begin{align}
\eta_{\bm{k}}^{3}&=\frac{1}{2}\left(L_{\bm{k}}+\sqrt{L_{\bm{k}}^{2}-4\left(\frac{P_{\bm{k}}^{2}}{9}+\frac{4}{3}R_{\bm{k}}\right)^{3}}\right),
\\
\zeta_{\bm{k}}^{3}&=\frac{1}{2}\left(L_{\bm{k}}-\sqrt{L_{\bm{k}}^{2}-4\left(\frac{P_{\bm{k}}^{2}}{9}+\frac{4}{3}R_{\bm{k}}\right)^{3}}\right),
\end{align}
\begin{equation}
L_{\bm{k}}=-\frac{2}{27}P_{\bm{k}}^{3}+\frac{8}{3}P_{\bm{k}}R_{\bm{k}}+Q_{\bm{k}}^{2},
\end{equation}
\begin{equation}
P_{\bm{k}}=2(\xi^{(0)})^{2}+(\xi_{\bm{k}}^{(x)})^{2}+(\xi_{\bm{k}}^{(y)})^{2}+(\xi_{-\bm{k}}^{(x)})^{2}+(\xi_{-\bm{k}}^{(y)})^{2}-2(\Delta^{(0)})^{2}-2(\Delta_{\bm{k}}^{(x)})^{2}-2|\Delta_{\bm{k}}^{(y)}|^{2},
\end{equation}
\begin{equation}
Q_{\bm{k}}=2\xi^{(0)}\{(\xi_{\bm{k}}^{(x)})^{2}+(\xi_{\bm{k}}^{(y)})^{2}-(\xi_{-\bm{k}}^{(x)})^{2}-(\xi_{-\bm{k}}^{(y)})^{2}\}-4\Delta^{(0)}\{(\xi_{\bm{k}}^{(x)}-\xi_{-\bm{k}}^{(x)})\Delta_{\bm{k}}^{(x)}-(\xi_{\bm{k}}^{(y)}+\xi_{-\bm{k}}^{(y)})\mathrm{Re}\Delta_{\bm{k}}^{(y)}\},
\end{equation}
\begin{align}
R_{\bm{k}}=&\{(\xi^{(0)})^{2}-(\Delta^{(0)})^{2}\}^{2}-(\xi^{(0)})^{2}\{(\xi_{\bm{k}}^{(x)})^{2}+(\xi_{\bm{k}}^{(y)})^{2}+(\xi_{-\bm{k}}^{(x)})^{2}+(\xi_{-\bm{k}}^{(y)})^{2}+2(\Delta_{\bm{k}}^{(x)})^{2}+2|\Delta_{\bm{k}}^{(y)}|^{2}\}
\nonumber \\
&-2(\Delta^{(0)})^{2}\{\xi_{\bm{k}}^{(x)}\xi_{-\bm{k}}^{(x)}-\xi_{\bm{k}}^{(y)}\xi_{-\bm{k}}^{(y)}+(\Delta_{\bm{k}}^{(x)})^{2}+(\mathrm{Re}\Delta_{\bm{k}}^{(y)})^{2}-(\mathrm{Im}\Delta_{\bm{k}}^{(y)})^{2}\}
\nonumber \\
&+4\xi^{(0)}\Delta^{(0)}\{(\xi_{\bm{k}}^{(x)}+\xi_{-\bm{k}}^{(x)})\Delta_{\bm{k}}^{(x)}-(\xi_{\bm{k}}^{(y)}-\xi_{-\bm{k}}^{(y)})\mathrm{Re}\Delta_{\bm{k}}^{(y)}\}
\nonumber \\
&+\{\xi_{\bm{k}}^{(x)}\xi_{-\bm{k}}^{(x)}+\xi_{\bm{k}}^{(y)}\xi_{-\bm{k}}^{(y)}-(\Delta_{\bm{k}}^{(x)})^{2}+|\Delta_{\bm{k}}^{(y)}|^{2}\}^{2}+(\xi_{-\bm{k}}^{(x)}\xi_{\bm{k}}^{(y)}-\xi_{\bm{k}}^{(x)}\xi_{-\bm{k}}^{(y)}+2\Delta_{\bm{k}}^{(x)}\mathrm{Re}\Delta_{\bm{k}}^{(y)})^{2}.
\end{align}
The unnormalized eigenvector $(U_{\pm,\mathrm{A}}(\bm{k}), U_{\pm,\mathrm{B}}(\bm{k}), V_{\pm,\mathrm{A}}(\bm{k}), V_{\pm,\mathrm{B}}(\bm{k}))$ is
\begin{eqnarray}
U_{\pm,A}(\bm{k})&=&\left\{(\xi_{-\bm{k}}^{(x)}+i\xi_{-\bm{k}}^{(y)})(\Delta_{\bm{k}}^{(x)}-i(\Delta_{\bm{k}}^{(y)})^{*})-(\xi^{(0)}+\omega_{\pm}(\bm{k}))\Delta^{(0)}\right\}\frac{V_{\pm,\mathrm{A}}(\bm{k})}{(\Delta^{(0)})^{2}-(\Delta_{\bm{k}}^{(x)})^{2}+\{(\Delta_{\bm{k}}^{(y)})^{*}\}^{2}} \nonumber \\
&&+\left\{(\xi^{(0)}+\omega_{\pm}(\bm{k}))(\Delta_{\bm{k}}^{(x)}-i(\Delta_{\bm{k}}^{(y)})^{*})-(\xi_{-\bm{k}}^{(x)}-i\xi_{-\bm{k}}^{(y)})\Delta^{(0)}\right\}\frac{V_{\pm,\mathrm{B}}(\bm{k})}{(\Delta^{(0)})^{2}-(\Delta_{\bm{k}}^{(x)})^{2}+\{(\Delta_{\bm{k}}^{(y)})^{*}\}^{2}},
\end{eqnarray}
\begin{eqnarray}
U_{\pm,\mathrm{B}}(\bm{k})&=&\left\{(\xi^{(0)}+\omega_{\pm}(\bm{k}))(\Delta_{\bm{k}}^{(x)}+i(\Delta_{\bm{k}}^{(y)})^{*})-(\xi_{-\bm{k}}^{(x)}+i\xi_{-\bm{k}}^{(y)})\Delta^{(0)}\right\}\frac{V_{\pm,\mathrm{A}}(\bm{k})}{(\Delta^{(0)})^{2}-(\Delta_{\bm{k}}^{(x)})^{2}+\{(\Delta_{\bm{k}}^{(y)})^{*}\}^{2}}
\nonumber \\
&&+\left\{(\xi_{-\bm{k}}^{(x)}-i\xi_{-\bm{k}}^{(y)})(\Delta_{\bm{k}}^{(x)}+i(\Delta_{\bm{k}}^{(y)})^{*})-(\xi^{(0)}+\omega_{\pm}(\bm{k}))\Delta^{(0)}\right\}\frac{V_{\pm,\mathrm{B}}(\bm{k})}{(\Delta^{(0)})^{2}-(\Delta_{\bm{k}}^{(x)})^{2}+\{(\Delta_{\bm{k}}^{(y)})^{*}\}^{2}},
\end{eqnarray}
\begin{eqnarray}
\frac{V_{\pm,\mathrm{A}}(\bm{k})}{(\Delta^{(0)})^{2}-(\Delta_{\bm{k}}^{(x)})^{2}+\{(\Delta_{\bm{k}}^{(y)})^{*}\}^{2}}&=&\{(\xi^{(0)})^{2}-(\omega_{\pm}(\bm{k}))^{2}\}(\Delta_{\bm{k}}^{(x)}-i(\Delta_{\bm{k}}^{(y)})^{*}))-\Delta^{(0)}(\xi_{\bm{k}}^{(x)}-\xi_{-\bm{k}}^{(x)}+i(\xi_{\bm{k}}^{(y)}+\xi_{-\bm{k}}^{(y)}))\omega_{\pm}(\bm{k})
\nonumber \\
&&+(\Delta_{\bm{k}}^{(x)}+i(\Delta_{\bm{k}}^{(y)})^{*})\{(\xi_{\bm{k}}^{(x)}+i\xi_{\bm{k}}^{(y)})(\xi_{-\bm{k}}^{(x)}-i\xi_{-\bm{k}}^{(y)})\}-(\Delta_{\bm{k}}^{(x)}-i\Delta_{\bm{k}}^{(y)})(\Delta_{\bm{k}}^{(x)}-i(\Delta_{\bm{k}}^{(y)})^{*})
 \nonumber \\
&&+(\Delta^{(0)})\left\{-\xi^{(0)}(\xi_{\bm{k}}^{(x)}+\xi_{-\bm{k}}^{(x)}+i(\xi_{\bm{k}}^{(y)}-\xi_{-\bm{k}}^{(y)}))+\Delta^{(0)}(\Delta_{\bm{k}}^{(x)}-i\Delta_{\bm{k}}^{(y)})\right\},
\nonumber \\
\end{eqnarray}
\begin{eqnarray}
\frac{V_{\pm,\mathrm{B}}(\bm{k})}{(\Delta^{(0)})^{2}-(\Delta_{\bm{k}}^{(x)})^{2}+\{(\Delta_{\bm{k}}^{(y)})^{*}\}^{2}}&=&-(\xi^{(0)}+\omega_{\pm}(\bm{k}))(\xi_{\bm{k}}^{(x)}+i\xi_{\bm{k}}^{(y)})(\Delta_{\bm{k}}^{(x)}+i(\Delta_{\bm{k}}^{(y)})^{*})
\nonumber \\
&&-(\xi^{(0)}-\omega_{\pm}(\bm{k}))(\xi_{-\bm{k}}^{(x)}+i\xi_{-\bm{k}}^{(y)})(\Delta_{\bm{k}}^{(x)}-i(\Delta_{\bm{k}}^{(y)})^{*})
\nonumber \\
&&+\Delta^{(0)}\left\{(\xi^{(0)})^{2}+(\xi_{\bm{k}}^{(x)}+i\xi_{\bm{k}}^{(y)})(\xi_{-\bm{k}}^{(x)}+i\xi_{-\bm{k}}^{(y)})-(\Delta^{(0)})^{2}\right.
\nonumber \\
&&\hspace{40pt}\left.+(\Delta_{\bm{k}}^{(x)})^{2}+((\Delta_{\bm{k}}^{(y)})^{*})^{2}-(\omega_{\pm}(\bm{k}))^{2}\right\}.
\end{eqnarray}
The normalized eigenvector $\bm{t}_{\pm}(\bm{k})$ can be written as
\begin{equation}
\bm{t}_{\pm}(\bm{k})=
\begin{pmatrix}
u_{\pm,A}(\bm{k})\\
u_{\pm,B}(\bm{k})\\
v_{\pm,A}(\bm{k})\\
v_{\pm,B}(\bm{k})
\end{pmatrix}
=
\begin{pmatrix}
\mathrm{e}^{i\rho_{\pm,1}(\bm{k})}\mathrm{cosh}\chi_{\pm}(\bm{k})\mathrm{cos}\mu_{\pm}(\bm{k}) \\
\mathrm{e}^{i\rho_{\pm,2}(\bm{k})}\mathrm{cosh}\chi_{\pm}(\bm{k})\mathrm{sin}\mu_{\pm}(\bm{k}) \\
\mathrm{e}^{i\rho_{\pm,3}(\bm{k})}\mathrm{sinh}\chi_{\pm}(\bm{k})\mathrm{cos}\nu_{\pm}(\bm{k}) \\
\mathrm{e}^{i\rho_{\pm,4}(\bm{k})}\mathrm{sinh}\chi_{\pm}(\bm{k})\mathrm{sin}\nu_{\pm}(\bm{k}) 
\end{pmatrix}
\end{equation}
where $\rho_{\pm,1}(\bm{k}),\cdots,\rho_{\pm,4}(\bm{k})\in\mathbb{R}$, $\mu_{\pm}(\bm{k}),\ \nu_{\pm}(\bm{k})\in[0,\pi/2]$ and $\chi_{\pm}(\bm{k})\geq0$.
These values satisfy
\begin{equation}
\mathrm{tanh}\chi_{\pm}(\bm{k})=\frac{\sqrt{|V_{\pm,A}(\bm{k})|^{2}+|V_{\pm,B}(\bm{k})|^{2}}}{\sqrt{|U_{\pm,A}(\bm{k})|^{2}+|U_{\pm,B}(\bm{k})|^{2}}}
\end{equation}
\begin{equation}
\mathrm{tan}\mu_{\pm}(\bm{k})=\left|\frac{U_{\pm,B}(\bm{k})}{U_{\pm,A}(\bm{k})}\right| \hspace{20pt} \mathrm{tan}\nu_{\pm}(\bm{k})=\left|\frac{V_{\pm,B}(\bm{k})}{V_{\pm,A}(\bm{k})}\right|
\end{equation}
\begin{equation}
\mathrm{e}^{i\rho_{\pm,1}(\bm{k})}=\frac{U_{\pm,A}(\bm{k})}{|U_{\pm,A}(\bm{k})|} \hspace{20pt} \mathrm{e}^{i\rho_{\pm,2}(\bm{k})}=\frac{U_{\pm,B}(\bm{k})}{|U_{\pm,B}(\bm{k})|} \hspace{20pt} \mathrm{e}^{i\rho_{\pm,3}(\bm{k})}=\frac{V_{\pm,A}(\bm{k})}{|V_{\pm,A}(\bm{k})|} \hspace{20pt} \mathrm{e}^{i\rho_{\pm,4}(\bm{k})}=\frac{V_{\pm,B}(\bm{k})}{|V_{\pm,B}(\bm{k})|}.
\end{equation}
In the case of $D_{\parallel}=0$ and $\phi_{\perp}=0$, we can get the simple expression of the magnon bands and the corresponding parameters of the eigenvectors as
\begin{eqnarray}
\omega_{1}(\bm{k})&=&\sqrt{(\xi^{(0)}+\xi_{\bm{k}}^{(x)})^{2}-(\Delta^{(0)}+\Delta_{\bm{k}}^{(x)})^{2}} \\
\omega_{2}(\bm{k})&=&\sqrt{(\xi^{(0)}-\xi_{\bm{k}}^{(x)})^{2}-(\Delta^{(0)}-\Delta_{\bm{k}}^{(x)})^{2}}
\end{eqnarray}
\begin{eqnarray}
\mathrm{tanh}\chi_{1}(\bm{k})&=&\sqrt{\frac{\xi^{(0)}+\xi_{\bm{k}}^{(x)}
-\omega_{1}(\bm{k})}{\xi^{(0)}+\xi_{\bm{k}}^{(x)}+\omega_{1}(\bm{k})}} \\
\mathrm{tanh}\chi_{2}(\bm{k})&=&\sqrt{\frac{\xi^{(0)}-\xi_{\bm{k}}^{(x)}
-\omega_{2}(\bm{k})}{\xi^{(0)}-\xi_{\bm{k}}^{(x)}+\omega_{2}(\bm{k})}}
\end{eqnarray}
\begin{equation}
\mathrm{tan}\mu_{1}(\bm{k})=\mathrm{tan}\nu_{1}(\bm{k})=\mathrm{tan}\mu_{2}(\bm{k})=\mathrm{tan}\nu_{2}(\bm{k})=1
\end{equation}
\begin{equation}
\mathrm{e}^{i\rho_{1,1}(\bm{k})}=\mathrm{e}^{i\rho_{1,2}(\bm{k})}=1 \hspace{20pt} \mathrm{e}^{i\rho_{1,3}(\bm{k})}=\mathrm{e}^{i\rho_{1,4}(\bm{k})}=-\mathrm{sgn}(\Delta^{(0)}+\Delta_{\bm{k}}^{(x)})
\end{equation}
\begin{equation}
\mathrm{e}^{i\rho_{2,1}(\bm{k})}=\mathrm{e}^{i\rho_{2,3}(\bm{k})}=1 \hspace{20pt} \mathrm{e}^{i\rho_{2,2}(\bm{k})}=\mathrm{e}^{i\rho_{2,4}(\bm{k})}=-1.
\end{equation}
The eigenvector of the edge modes is given by
\begin{align}
\bm{t}_{\mu}(k_{\bm{\delta}_{1}})=&
(u_{\mu,\mathrm{A}}^{(1)}(k_{\bm{\delta}_{1}}),u_{\mu,\mathrm{B}}^{(1)}(k_{\bm{\delta}_{1}}),\cdots,u_{\mu,\mathrm{A}}^{(2L)}(k_{\bm{\delta}_{1}}),u_{\mu,\mathrm{B}}^{(2L)}(k_{\bm{\delta}_{1}}),
\nonumber\\
&\hspace{10pt}v_{\mu,\mathrm{A}}^{(1)}(k_{\bm{\delta}_{1}}),v_{\mu,\mathrm{B}}^{(1)}(k_{\bm{\delta}_{1}}),\cdots,v_{\mu,\mathrm{A}}^{(2L)}(k_{\bm{\delta}_{1}}),v_{\mu,\mathrm{B}}^{(2L)}(k_{\bm{\delta}_{1}}))
,
\end{align}
where $\mu=\mathrm{L},\mathrm{R}$ denotes the left and right edge modes.
The local density of states of the two edge modes are given by
\begin{equation}
d_{\mu}^{(j)}(k_{\bm{\delta}_{1}})=
\sum_{\alpha=\mathrm{A},\mathrm{B}}\left\{
|u_{\mu,\alpha}^{(j)}(k_{\bm{\delta}_{1}})|^{2}
+|v_{\mu,\alpha}^{(j)}(k_{\bm{\delta}_{1}})|^{2}
\right\}
,
\end{equation}
where $j$ is the site index along the $\bm{\delta}_{2}$-direction.

%*%*%*%*%*%*%*%*%*%*%*%*%*%*%*%*%*%*%*%*%*%*%*%*%*%*%*%*%*%*%*%*%*
\section{Direction of the nonreciprocal propagation}
\label{app3}
Using Eq.(\ref{eq:egeq}) and the relation, $\bm{t}_{\pm}^{\dagger}\Sigma^{z}\bm{t}_{\pm}(\bm{k})=1$, the derivative of $\omega_{\pm}(\bm{k})$ with respect to $\bm{k}$ is given as follows;
\begin{align}
\bm{\nabla}_{\bm{k}}\omega_{\pm}(\bm{k})&=
\bm{\nabla}_{\bm{k}}[\bm{t}_{\pm}^{\dagger}(\bm{k})H_{\mathrm{BdG}}(\bm{k})\bm{t}_{\pm}(\bm{k})]
\nonumber \\
&=\bm{t}_{\pm}^{\dagger}(\bm{k})[\bm{\nabla}_{\bm{k}}H_{\mathrm{BdG}}(\bm{k})]\bm{t}_{\pm}(\bm{k})+\omega_{\pm}(\bm{k})\bm{\nabla}_{\bm{k}}[\bm{t}_{\pm}^{\dagger}(\bm{k})\Sigma^{z}\bm{t}_{\pm}(\bm{k})]
\nonumber \\
&=\bm{t}_{\pm}^{\dagger}(\bm{k})[\bm{\nabla}_{\bm{k}}H_{\mathrm{BdG}}(\bm{k})]\bm{t}_{\pm}(\bm{k})
.
\end{align}
In the case of $\phi_{\perp}=0$, we find
\begin{equation}
\bm{\nabla}_{\bm{k}}\omega_{\pm}(\bm{k})|_{\bm{k}=\bm{0}}=
\pm\mathrm{sgn}(\xi^{(0)}\xi_{\bm{0}}^{(x)}-\Delta^{(0)}\Delta_{\bm{0}}^{(x)})\sqrt{2}D_{\parallel}S\ \mathrm{sin}\theta_{\parallel}(\bm{e}_{x}\mathrm{cos}\phi_{\parallel}-\bm{e}_{y}\mathrm{sin}\phi_{\parallel}).
\end{equation}
Thus, the two bands have velocities of different sign at $\Gamma$ point, 
and $D_{\parallel}$ pushes the minima of these bands in the opposite directions, making the bands nonreciprocal.

%*%*%*%*%*%*%*%*%*%*%*%*%*%*%*%*%*%*%*%*%*%*%*%*%*%*%*%*%*%*%*%*%*

%*%*%*%*%*%*%*%*%*%*%*%*%*%*%*%*%*%*
%*%*%*%*%*%*%*%*%*%*%*%*%*%*%*%*%*%*
\begin{figure*}[tb]
\includegraphics[width=17cm]{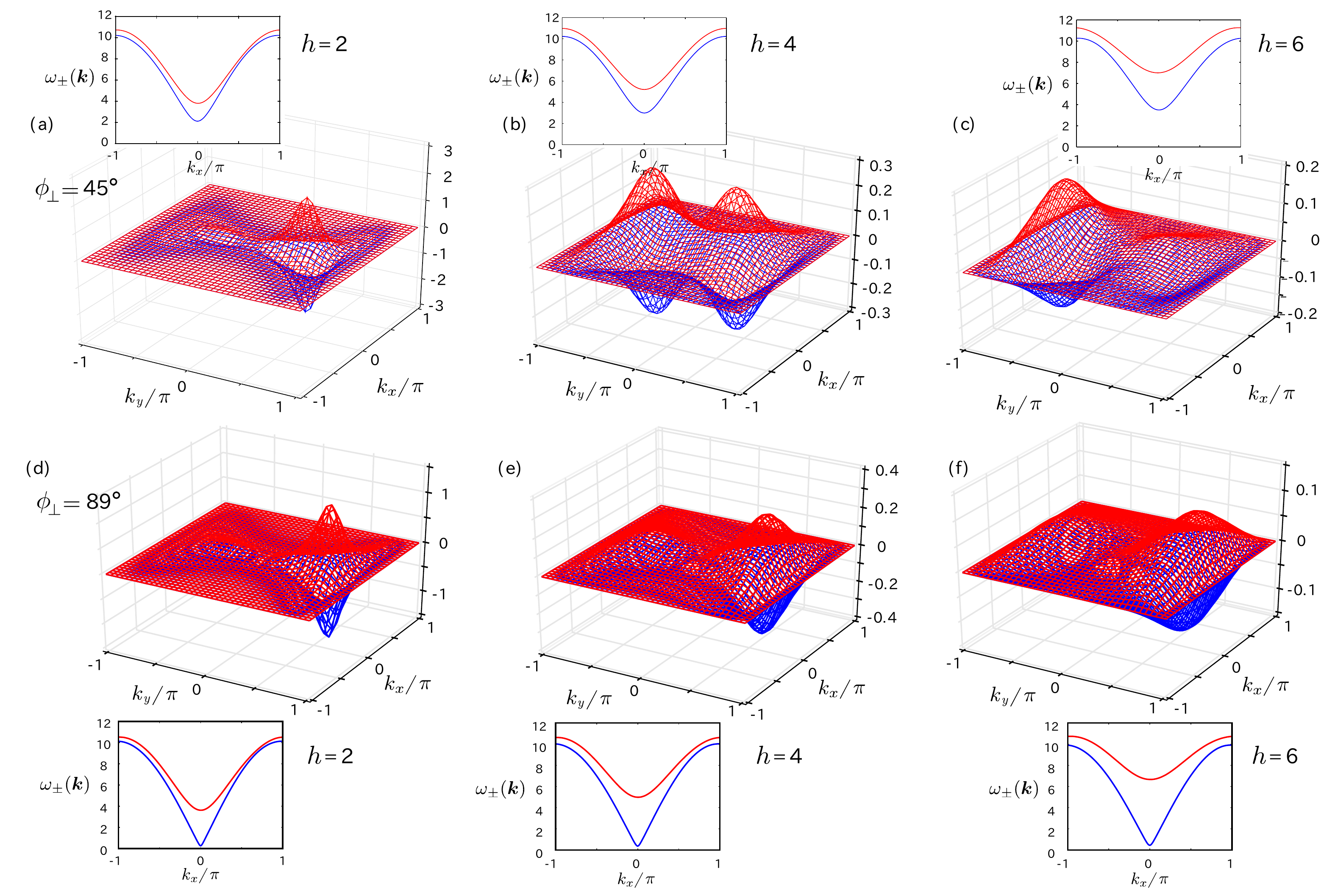}
\caption{ 
Details of the Berry curvature $\Omega_{xy}^{(\pm)}(\bm k)$ for 
field angles (a)-(c) $\phi_\perp=45^\circ$  and (d)-(f) $\phi_\perp=89^\circ$, with $h/J=$2,4, and 6. 
The other parameters are taken as the same as Fig.\ref{f3}, 
$J=1.0$, $D_\parallel = D_\perp = 0.2$, $\Lambda= 0.05$.
When $\phi_\perp=45^\circ$, the peak position shifts with increasing $h$ in the $-k_x$ direction along $k_y=0$. 
For $\phi_\perp=89^\circ$, such peak-shift is not observed. 
The corresponding magnon bands along the $k_y=0$ line are shown together. 
}
\label{fapp3}
\end{figure*}
%*%*%*%*%*%*%*%*%*%*%*%*%*%*%*%*%*%*
\section{Development of the Berry curvature with magnetic field}
\label{app4}
We show in Fig.\ref{fapp3} the field angle dependence of the Berry curvature.
In the case of $\phi_{\perp}=45^{\circ}$,
the peak position shifts to the $-k_{x}$ direction along $k_{y}=0$ line due to the in-plane component of the magnetic field, $\bm{h}_{\parallel}$.
While in the case of $\phi_{\perp}=89^{\circ}$,
the peak position does not shift.
This is because the in-plane component of the magnetic field, $|\bm{h}_{\parallel}|=h\mathrm{cos}89^{\circ}$, is very small.
\newpage

\end{widetext}

%*%*%*%*%*%*%*%*%*%*%*%*%*%*%*%*%*
%*%*%*%*%*%*%*%*%*%*%*%*%*%*%*%*%*
%%%%%%%%%  REFER  %%%%%%%%%%%%%%%%%%%%%%%%%%%%%%%%%%%%%%%

%%%%%%%%%%%%%%%%%%%%%%%%%%%%%%%%%%%%%%%%%%%%%%%%%%%%%%%%%%%%%%%%%%%%
%\end{multicols}

\begin{references}
\bibitem{honey}Only in specific cases, the cancellation of the effect of the U(1) gauge field is avoided, such as the system with spatial dependent electric field\cite{nakata17F,nakata17AF} or having a nontrivial spin texture in real space\cite{hoogdalem13}.
\end{references}
\end{document}